\documentclass[11pt]{article}

\usepackage[usenames,dvipsnames,table]{xcolor}
\usepackage[utf8]{inputenc}

\pdfoutput=1 
\usepackage{jheppub} 

\usepackage{graphicx}
\usepackage{amsmath, amssymb}

\newcommand{\be}{\begin{eqnarray}}
\newcommand{\ee}{\end{eqnarray}}
\newcommand{\bea}{\begin{eqnarray}}
\newcommand{\eea}{\end{eqnarray}}

\newcommand{\bn}{\begin{enumerate}}
\newcommand{\en}{\end{enumerate}}

\def\half{\frac{1}{2}}

\title{Non minimal D-type conformal matter compactified on three punctured spheres}

\preprint{}

\author{Evyatar Sabag}

\affiliation{Department of Physics, Technion, Haifa, 32000, Israel}

\emailAdd{sevyatar@campus.technion.ac.il}

\abstract{We study compactifications of $6d$ non minimal $(D_{p+3},D_{p+3})$ type conformal matter. These can be described by $N$ M5-branes probing a $D_{p+3}$-type singularity. We derive $4d$ Lagrangians corresponding to compactifications of such $6d$ SCFTs on three punctured spheres (trinions) with two maximal punctures and one minimal puncture.
The trinion models are described by simple $\mathcal{N}=1$ quivers with $SU(2N)$ gauge nodes. We derive the trinion Lagrangians using RG flows between the aforementioned $6d$ SCFTs with different values of $p$ and their relations to matching RG flows in their compactifications to $4d$.
The suggested trinions are shown to reduce to known models in the minimal case of $N=1$. Additional checks are made to show the new minimal punctures uphold the expected S-duality  between models in which we exchange two such punctures. We also show that closing the new minimal puncture leads to expected flux tube models. 
}

\begin{document} 

\maketitle
\flushbottom

\section{Introduction}

In the recent decade, following the influential work of \cite{Gaiotto:2009we} there has been a plethora of work relating compactifications of $6d$ SCFTs on a Riemann surface to Lagrangian $4d$ SCFTs. Such relations were understood for general Riemann surfaces with fluxes for several specific $6d$ SCFTs, including $A_1$ and $A_2$ $(2,0)$ \cite{Gaiotto:2009we,Gadde:2015xta}, $A_1$ $(2,0)$ probing a $\mathbb{Z}_2$ singularity \cite{Gaiotto:2015usa,Razamat:2016dpl,Hanany:2015pfa}, the rank one E-string \cite{Kim:2017toz,Razamat:2019vfd}, and for $SO(8)$ and $SU(3)$ minimal SCFTs \cite{Razamat:2018gro}. In addition, recently such relations have been found for entire classes of $6d$ SCFTs for general Riemann surfaces with fluxes, including $A_1$ $(2,0)$ probing a $\mathbb{Z}_k$ singularity and $A_0$ $(2,0)$ probing a $D_{N+3}$ singularity \cite{Razamat:2019ukg,Razamat:2020bix}. There are many more SCFTs for which only special surfaces Lagrangians are known. For example $(2,0)$ SCFTs probing higher rank E-string or ADE singularities on tori surfaces or genus zero surfaces with two punctures or less \cite{Bah:2017gph,Kim:2018bpg,Kim:2018lfo,Chen:2019njf,Pasquetti:2019hxf}. The relations between $4d$ and $6d$ SCFTs lead to many new understandings regarding dualities and their relation to geometry, as well as emergent IR symmetries \cite{Razamat:2018gbu}.

One method to find $4d$ Lagrangians related to $6d$ compactifications is by using anomaly predictions from $6d$ \cite{Razamat:2019vfd}. These are used to predict the number of vector and hyper multiplets of the $4d$ theory assuming it is a conformal gauge theory with all the gauge couplings having vanishing one loop $\beta$ functions. In many such cases the possibilities are very limited and one can find a quiver description that matches the requirements. With the quiver at hand various checks can be performed to verify the result is as expected. Such strategies were used in \cite{Razamat:2019vfd,Razamat:2020gcc,Razamat:2020bix}, and the classification in \cite{Razamat:2020pra} can make such efforts much simpler. In addition, similar methods were used in \cite{Zafrir:2019hps}.

Another strategy employed to find $4d$ Lagrangians compactified from $6d$ SCFTs takes advantage of $5d$ domain walls. This strategy can be used in cases where compactifying the $6d$ SCFT on a circle, with or without holonomies and twists, results in an effective $5d$ gauge theory. An additional segment or circle compactification reduces the theory to $4d$. This $4d$ theory can be obtained from $5d$ duality domain walls \cite{Gaiotto:2014ina,Kim:2017toz,Kim:2018bpg,Kim:2018lfo}. On each side of the domain wall there is an effective $5d$ theory, both theories are obtained from the same $6d$ SCFT compactified on a circle but with different values of holonomies. These nontrivial holonomies lead to nontrivial flux on the Riemann surface related to the $4d$ model \cite{Chan:2000qc,Kim:2017toz}. This method leads to tori and tube (sphere with two punctures) related $4d$ Lagrangians.

The former strategy can be supplemented with a method to construct $4d$ Lagrangians for compactifications with additional punctures that was shown in \cite{Razamat:2019mdt}. In this approach one needs to consider both a flux compactification of $6d$ SCFTs and a flow induced by triggering a vacuum expectation value (vev) to certain $6d$ operators. It was found that first compactifying to $4d$ on a Riemann surface with flux and then setting the vev inducing a flow, is equivalent to first setting the vev and flowing to a new $6d$ SCFT and then compactifying on a different Riemann surface. The latter surface differs from the former in flux and possibly has additional punctures depending on the former surface flux. This was shown to be explicitly true by examining known class $\mathcal{S}_k$ Lagrangians, and also checked for certain known index limits \cite{Razamat:2018zus} of class $\mathcal{S}_k$ models with no known Lagrangian. Class $\mathcal{S}_k$ models are obtained by flux compactifying on a Riemann surface the $6d$ $(1,0)$ SCFTs described by a stack of M5-branes probing a $\mathbb{Z}_k$ singularity. This method was later successfully used to construct new unknown $4d$ Lagrangians resulting from compactifications of $6d$ $(1,0)$ SCFTs described by a single M5-brane probing a $D_{N+3}$ singularity \cite{Razamat:2019ukg}.

In this note we will apply the procedure of generating $4d$ models described by a Riemann surface with extra punctures on the non minimal $(D_{p+3}, D_{p+3})$ conformal matter $6d$ models \cite{DelZotto:2014hpa}. We will also denote these models in abbreviation as the $D_{p+3}$ SCFTs. These models are $6d$ $(1,0)$ SCFTs residing on a stack of M5-branes probing a $D_{p+3}$ singularity, and are sometimes denoted by $\mathcal{T}\left(SO(2p+6),N\right)$ \cite{DelZotto:2014hpa}. The case of $N=1$ known as the $(D_{p+3}, D_{p+3})$ minimal conformal matter was studied and mapped thoroughly in \cite{Razamat:2019ukg,Razamat:2020bix} and we will use these results for consistency checks.

The aforementioned duality domain wall approach was already successfully used for the $\mathcal{T}\left(SO(2p+6),N\right)$ models to find $4d$ theories corresponding to spheres with two maximal punctures (tubes) \cite{Kim:2018lfo}. However theories corresponding to Riemann surfaces with more than two punctures are unknown for $N>1$. Here we use flows between $\mathcal{T}\left(SO(2p+6),N\right)$ SCFTs with different $p$ to derive $4d$ models corresponding to spheres with two maximal punctures, with $SU(N)^4 \times SU(2N)^{p}$ symmetry, and one minimal puncture with a $U(1)$ symmetry. The obtained three punctured models are quiver theories of $SU(2N)$ gauge nodes with $8N$ flavors. These models are then verified to be consistent with known results and expected dualities. In addition we check for consistency when we close the new minimal puncture and recover a tube theory and also match anomalies to the ones predicted from $6d$.

This paper is organized as follows. In Section \ref{S:Trinion} we present the main result of the $4d$ Lagrangian corresponding to a three punctured sphere (trinion) compactification of $\mathcal{T}\left(SO(2p+6),N\right)$, and show it is consistent under all the checks performed. In Section \ref{S:Derivation} we show the derivation of the Lagrangians discussed in the former section using $6d$ and $4d$ RG-flows. In addition, there are Several appendices with information on notations and some additional derivations. 

\section{Trinion compactification of the $D_{p+3}$ SCFT}\label{S:Trinion}
In this section we propose a $4d$ Lagrangian for a three punctured sphere with two maximal and one minimal puncture, of the $D_{p+3}$ conformal matter SCFT composed of $N$ copies of the $D_{p+3}$ minimal conformal matter SCFT. The derivation of this result using RG flows is carried out in the next section. Here we will focus on the trinion properties, and give evidence that the claimed new minimal puncture indeed upholds puncture properties, and is also consistent with known results from \cite{Razamat:2019ukg}.

Let us note that all the compactifications of $D_{p+3}$ minimal conformal matter SCFT ($N=1$) on a Riemann surface can be constructed using the Lagrangians found in \cite{Razamat:2019ukg}. As for the case of $N>1$ only flux tubes/tori compactifications are known from \cite{Kim:2018lfo}. We will use these previous results to preform consistency checks for our new models.

\subsection{The trinion}

\begin{figure}[t]
	\centering
  	\includegraphics[scale=0.315]{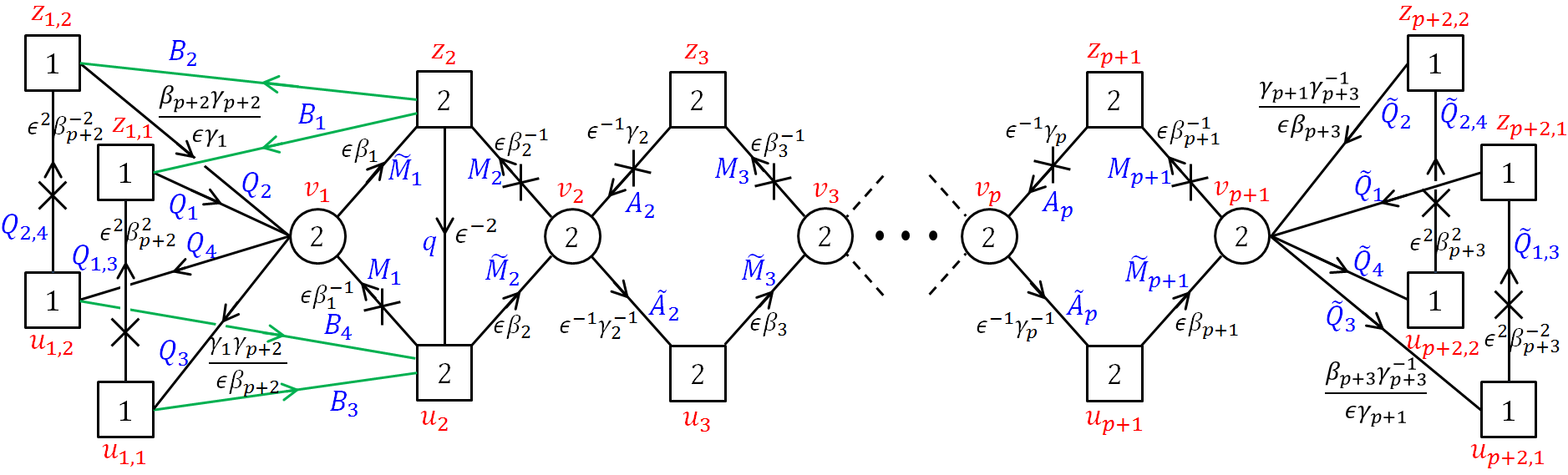}
    \caption{A quiver diagram of a trinion with two maximal punctures with $SU(N)^4 \times SU(2N)^p$ symmetry and one minimal puncture with a $U(1)_\epsilon$ symmetry for the $D_{p+3}$ conformal matter SCFT with general $N$. The squares and circles denote $SU(nN)$ global and gauge symmetries, respectively, where $n$ is the number inside the shapes. The fields transforming under the gauge symmetry have R-charge $1/2$, the gauge singlet fields not marked by $X$ have R-charge $1$. Flip fields are marked by $X$'s on bifundamental fields, they are coupled through the superpotential to baryonic operators built from these bifundamentals and have R-charge $2-N$. In addition there are six more flip fields that need to be included flipping each of the baryonic operators $Q_i^N Q_4^N$ and $\widetilde{Q}_i^N \widetilde{Q}_j^N$, with $i,j=1,2,3$ and $i>j$. In addition, each closed loop of fields has a superpotential term turned on for it. $\epsilon$ is the fugacity associated to the additional minimal puncture, while $\beta_i$ and $\gamma_j$ are related to the internal symmetries that arise from $6d$. In blue we write the field names, in red we write the non-abelian symmetry associated fugacities, and in black we write the charges of each field in terms of the fugacities associated with the different $U(1)$'s.}
    \label{F:DTrinion}
\end{figure}

The $D_{p+3}$ trinion quiver is shown in Figure  \ref{F:DTrinion}. It has two maximal punctures with $SU(N)^4 \times SU(2N)^p$ global symmetries and a third minimal puncture with $U(1)_{\epsilon}$ symmetry. The minimal puncture symmetry doesn't enhance in the general case of $N>1$ unlike the case for $N=1$ where it gets enhanced to $SU(2)$ \cite{Razamat:2019ukg}. The theory has the following superpotential,
\be
W & = & (M_1 \widetilde{M}_1 + M_2 \widetilde{M}_2)q + \sum_{i=2}^{p} A_{i} \widetilde{A}_{i} M_{i+1} \widetilde{M}_{i+1} + \sum_{i=1}^2 B_i Q_i \widetilde{M}_1 + \sum_{j=3}^4 B_j Q_j M_1\nonumber\\
 & & + Q_{1,3} Q_1 Q_3 + Q_{2,4} Q_2 Q_4 + \widetilde{Q}_{1,3} \widetilde{Q}_1 \widetilde{Q}_3 + \widetilde{Q}_{2,4} \widetilde{Q}_2 \widetilde{Q}_4\nonumber\\
 & &  + \sum_{i=1}^{p+1} F_i M_i^{2N} + \sum_{j=2}^{p} \widetilde{F}_j A_j^{2N} + F_{1,3} Q_{1,3}^{N} + F_{2,4} Q_{2,4}^{N} + \widetilde{F}_{1,3} \widetilde{Q}_{1,3}^{N} + \widetilde{F}_{2,4} \widetilde{Q}_{2,4}^{N}\nonumber\\
 & & + F_{14} Q_1^N Q_4^N + F_{24} Q_2^N Q_4^N + F_{34} Q_3^N Q_4^N + \widetilde{F}_{12} \widetilde{Q}_1^N \widetilde{Q}_2^N + \widetilde{F}_{13} \widetilde{Q}_1^N \widetilde{Q}_3^N + \widetilde{F}_{23} \widetilde{Q}_2^N \widetilde{Q}_3^N ,\nonumber\\
\ee
where we suppressed the $SU(N)$ and $SU(2N)$ indices for brevity as they contract in a trivial manner. The different field names appear in Figure \ref{F:DTrinion}.
The fields denoted by $F$ are gauge singlet flip fields, needed for consistency with the known trinons of the $N=1$ cases. 

Arranging the fields, superpotential, gauge and global symmetries information into one expression can be done using the superconformal index \cite{Kinney:2005ej,Romelsberger:2005eg,Dolan:2008qi,Rastelli:2016tbz} displayed here,\footnote{See Appendix \ref{A:indexdefinitions} for index definitions and notations.}
\be
\label{E:DTrinion}
\mathcal{I}_{\boldsymbol{z},\boldsymbol{u},\epsilon}^{T(N,p)} & = & \left(\frac{\kappa^{2N-1}}{(2N)!}\right)^{p+1}\prod_{i=1}^{2N-1}\prod_{a=1}^{p+1}\oint\frac{dv_{a}^{(i)}}{2\pi iv_{a}^{(i)}}\frac{1}{\prod_{i\ne j}^{2N}\prod_{a=1}^{p+1}\Gamma_{e}\left(v_{a}^{(i)}\left(v_{a}^{(j)}\right)^{-1}\right)}\times\nonumber\\
(2) & & \prod_{j=1}^{2N}\prod_{I=1}^{N}\Gamma_{e}\left(\left(pq\right)^{\frac{1}{2}}\frac{\beta_{p+2}\gamma_{1}\gamma_{p+2}}{\beta_{1}}\frac{z_{1,1}^{(I)}}{z_{2}^{(j)}}\right)\Gamma_{e}\left(\left(pq\right)^{\frac{1}{2}}\frac{\gamma_{1}}{\beta_{1}\beta_{p+2}\gamma_{p+2}}\frac{z_{1,2}^{(I)}}{z_{2}^{(j)}}\right)\times\nonumber\\
(3)&&\prod_{j=1}^{2N}\prod_{I=1}^{N}\Gamma_{e}\left(\left(pq\right)^{\frac{1}{2}}\frac{\beta_{1}\beta_{p+2}}{\gamma_{1}\gamma_{p+2}}\frac{u_{2}^{(j)}}{u_{1,1}^{(I)}}\right)\Gamma_{e}\left(\left(pq\right)^{\frac{1}{2}}\frac{\beta_{1}\gamma_{p+2}}{\beta_{p+2}\gamma_{1}}\frac{u_{2}^{(j)}}{u_{1,2}^{(I)}}\right)\times\nonumber\\
(4)&&\Gamma_{e}\left(\left(pq\right)^{1-\frac{N}{2}}\frac{\beta_{p+2}^{\pm2N}}{\epsilon^{2N}}\right)\prod_{I,J=1}^{N}\Gamma_{e}\left(\left(pq\right)^{\frac{1}{2}}\epsilon^{2}\beta_{p+2}^{2}\frac{z_{1,1}^{(I)}}{u_{1,1}^{(J)}}\right)\Gamma_{e}\left(\left(pq\right)^{\frac{1}{2}}\frac{\epsilon^{2}}{\beta_{p+2}^{2}}\frac{z_{1,2}^{(I)}}{u_{1,2}^{(J)}}\right)\times\nonumber\\
(5)&&\prod_{j=1}^{2N}\prod_{I=1}^{N}\Gamma_{e}\left(\left(pq\right)^{\frac{1}{4}}\frac{1}{\epsilon\beta_{p+2}\gamma_{1}\gamma_{p+2}}\frac{v_{1}^{(j)}}{z_{1,1}^{(I)}}\right)\Gamma_{e}\left(\left(pq\right)^{\frac{1}{4}}\frac{\beta_{p+2}\gamma_{p+2}}{\epsilon\gamma_{1}}\frac{v_{1}^{(j)}}{z_{1,2}^{(I)}}\right)\times\nonumber\\
(6)&&\prod_{j=1}^{2N}\prod_{I=1}^{N}\Gamma_{e}\left(\left(pq\right)^{\frac{1}{4}}\frac{\gamma_{1}\gamma_{p+2}}{\epsilon\beta_{p+2}}\frac{u_{1,1}^{(I)}}{v_{1}^{(j)}}\right)\Gamma_{e}\left(\left(pq\right)^{\frac{1}{4}}\frac{\beta_{p+2}\gamma_{1}}{\epsilon\gamma_{p+2}}\frac{u_{1,2}^{(I)}}{v_{1}^{(j)}}\right)\times\nonumber\\
(7)&&\Gamma_{e}\left(\left(pq\right)^{1-\frac{N}{2}}\frac{\beta_{1}^{2N}}{\epsilon^{2N}}\right)\prod_{i,j=1}^{2N}\Gamma_{e}\left(\left(pq\right)^{\frac{1}{4}}\epsilon\beta_{1}\frac{z_{2}^{(j)}}{v_{1}^{(i)}}\right)\Gamma_{e}\left(\left(pq\right)^{\frac{1}{4}}\frac{\epsilon}{\beta_{1}}\frac{v_{1}^{(i)}}{u_{2}^{(j)}}\right)\Gamma_{e}\left(\left(pq\right)^{\frac{1}{2}}\frac{1}{\epsilon^{2}}\frac{u_{2}^{(i)}}{z_{2}^{(j)}}\right)\nonumber\\
(8)&&\Gamma_{e}\left(\left(pq\right)^{1-\frac{N}{2}}\frac{\beta_{2}^{2N}}{\epsilon^{2N}}\right)\prod_{i,j=1}^{2N}\Gamma_{e}\left(\left(pq\right)^{\frac{1}{4}}\frac{\epsilon}{\beta_{2}}\frac{z_{2}^{(i)}}{v_{2}^{(j)}}\right)\Gamma_{e}\left(\left(pq\right)^{\frac{1}{4}}\epsilon\beta_{2}\frac{v_{2}^{(j)}}{u_{2}^{(i)}}\right)\times\nonumber\\
(9)&&\prod_{a=2}^{p}\underline{\Gamma_{e}\left(\left(pq\right)^{1-\frac{N}{2}}\frac{\epsilon^{2N}}{\gamma_{a}^{2N}}\right)}\prod_{i,j=1}^{2N}\Gamma_{e}\left(\left(pq\right)^{\frac{1}{4}}\frac{\gamma_{a}}{\epsilon}\frac{v_{a}^{(i)}}{z_{a+1}^{(j)}}\right)\Gamma_{e}\left(\left(pq\right)^{\frac{1}{4}}\frac{1}{\epsilon\gamma_{a}}\frac{u_{a+1}^{(j)}}{v_{a}^{(i)}}\right)\times\nonumber\\
(10)&&\prod_{b=3}^{p+1}\Gamma_{e}\left(\left(pq\right)^{1-\frac{N}{2}}\frac{\beta_{b}^{2N}}{\epsilon^{2N}}\right)\prod_{i,j=1}^{2N}\Gamma_{e}\left(\left(pq\right)^{\frac{1}{4}}\frac{\epsilon}{\beta_{b}}\frac{z_{b}^{(i)}}{v_{b}^{(j)}}\right)\Gamma_{e}\left(\left(pq\right)^{\frac{1}{4}}\epsilon\beta_{b}\frac{v_{b}^{(j)}}{u_{b}^{(i)}}\right)\times\nonumber\\
(11)&&\prod_{i=1}^{2N}\prod_{J=1}^{N}\Gamma_{e}\left(\left(pq\right)^{\frac{1}{4}}\frac{\gamma_{p+1}}{\epsilon\beta_{p+3}\gamma_{p+3}}\frac{v_{p+1}^{(i)}}{z_{p+2,2}^{(J)}}\right)\Gamma_{e}\left(\left(pq\right)^{\frac{1}{4}}\frac{\beta_{p+3}\gamma_{p+1}\gamma_{p+3}}{\epsilon}\frac{v_{p+1}^{(i)}}{z_{p+2,1}^{(J)}}\right)\times\nonumber\\
(12)&&\prod_{i=1}^{2N}\prod_{J=1}^{N}\Gamma_{e}\left(\left(pq\right)^{\frac{1}{4}}\frac{\beta_{p+3}}{\epsilon\gamma_{p+1}\gamma_{p+3}}\frac{u_{p+2,1}^{(J)}}{v_{p+1}^{(i)}}\right)\Gamma_{e}\left(\left(pq\right)^{\frac{1}{4}}\frac{\gamma_{p+3}}{\epsilon\beta_{p+3}\gamma_{p+1}}\frac{u_{p+2,2}^{(J)}}{v_{p+1}^{(i)}}\right)\times\nonumber\\
(13)&&\Gamma_{e}\left(\left(pq\right)^{1-\frac{N}{2}}\frac{\beta_{p+3}^{\pm2N}}{\epsilon^{2N}}\right)\prod_{I,J=1}^{N}\Gamma_{e}\left(\left(pq\right)^{\frac{1}{2}}\frac{\epsilon^{2}}{\beta_{p+3}^{2}}\frac{z_{p+2,1}^{(I)}}{u_{p+2,1}^{(J)}}\right)\Gamma_{e}\left(\left(pq\right)^{\frac{1}{2}}\epsilon^{2}\beta_{p+3}^{2}\frac{z_{p+2,2}^{(I)}}{u_{p+2,2}^{(J)}}\right)\times\nonumber\\
(14)&&\underline{\Gamma_{e}\left(\left(pq\right)^{1-N/2}\epsilon^{2N}\gamma_{p+2}^{2N}\right)\Gamma_{e}\left(\left(pq\right)^{1-N/2}\epsilon^{2N}\beta_{p+2}^{-2N}\right)\Gamma_{e}\left(\left(pq\right)^{1-N/2}\epsilon^{2N}\gamma_{1}^{-2N}\right)}\times\nonumber\\
(15)&&\underline{\Gamma_{e}\left(\left(pq\right)^{1-N/2}\epsilon^{2N}\gamma_{p+3}^{2N}\right)\Gamma_{e}\left(\left(pq\right)^{1-N/2}\epsilon^{2N}\beta_{p+3}^{-2N}\right)\Gamma_{e}\left(\left(pq\right)^{1-N/2}\epsilon^{2N}\gamma_{p+1}^{-2N}\right)}\, .
\ee
The index fields are arranged beginning to end in the order they appear from left to right in Figure \ref{F:DTrinion}. In addition we have added the line numbers to the left of the expression for clarity. In the first line we write the contribution and integration on the gauge fields. Lines two and three show the $B_i$ fields. Lines $4-6$ present the $Q$ fields and the flipping fields $F_{1,3}$ and $F_{2,4}$, while lines $11-13$ show the respective tilted fields. Lines $7-8$ display the fields and flippings of the left rectangle in Figure \ref{F:DTrinion}, while lines $9-10$ show the fields of the rest of the rectangles. The last two lines present the additional flipping fields not appearing in Figure \ref{F:DTrinion}. These flipping fields and the other underlined fields are ones added for consistency with the known $N=1$ case found in \cite{Razamat:2019ukg}. 

Now, with the trinion at hand we want to specify the maximal punctures properties and how they can be glued to one another. First we note the operators in the fundamental representation of the punctures symmetry. For the $\textbf{u}$ maximal puncture with associated symmetry $SU(N)^4 \times SU(2N)^p$ the operators are $B_{j+2}$ in the bifundamental of $SU(N)_{u_{1,j}}$ and $SU(2N)_{u_2}$, and $\widetilde{M}_{p+1}\widetilde{Q}_{j+2}$ in the bifundamental of $SU(2N)_{u_{p+1}}$ and $SU(N)_{u_{p+2,j}}$ with $j=1,2$. In addition there are the operators $\widetilde{M}_{i}\widetilde{A}_{i}$ in the bifundamental of $SU(2N)_{u_{i}}$ and $SU(2N)_{u_{i+1}}$ with $i=2,...,p$. For the $\textbf{z}$ maximal puncture with associated symmetry $SU(N)^4 \times SU(2N)^p$ the operators are $B_{j}$ in the bifundamental of $SU(N)_{z_{1,j}}$ and $SU(2N)_{z_2}$, and $M_{p+1}\widetilde{Q}_{j}$ in the bifundamental of $SU(2N)_{z_{p+1}}$ and $SU(N)_{z_{p+2,j}}$ with $j=1,2$. These are joined by the operators $M_{i} A_{i}$ in the bifundamental of $SU(2N)_{z_{i}}$ and $SU(2N)_{z_{i+1}}$ with $i=2,...,p$.\footnote{Note that the $B_i$ operators were added to get punctures coming from boundary conditions $(+,+,...,+)$ and $(-,-,...,-)$ for the $\textbf{z}$ and $\textbf{u}$ punctures using the language of \cite{Kim:2018lfo}. The $B_i$ flip the sign of the first two entries out of $p+3$ in the boundary conditions, and without them the punctures would be of "type" $(-,-,+,...,+)$ and $(+,+,-,...,-)$ for the $\textbf{z}$ and $\textbf{u}$.}
We refer to these collections of operators as ``moment maps'' by abuse of terminology and denote them as $\widehat{M}^{(X)}$ with $X$ standing for the type of puncture. 
Thus, the ``moment maps'' for the maximal punctures are
\be
\label{E:MinDNMomentMaps}
\widehat M^{(\textbf{u})} &:&\ \,  \{\widehat M^{(u_{1,1},u_2)}: \beta_1 \beta_{p+2} \gamma_1^{-1} \gamma_{p+2}^{-1},\; \widehat M^{(u_{1,2},u_2)}: \beta_1 \beta_{p+2}^{-1} \gamma_1^{-1} \gamma_{p+2},\; \{ \widehat M^{(u_j,u_{j+1})}: \{ \beta_{j} \gamma_j^{-1}\} \}_{j=2}^{p}, \nonumber\\
 & & \ \,  \widehat M^{(u_{p+1},u_{p+2,1})}: \beta_{p+1} \beta_{p+3} \gamma_{p+1}^{-1} \gamma_{p+3}^{-1},\; \widehat M^{(u_{p+1},u_{p+2,2})}: \beta_{p+1} \beta_{p+3}^{-1} \gamma_{p+1}^{-1} \gamma_{p+3}\}\,,\nonumber\\
\widehat M^{(\textbf{z})} &:&\ \,  \{\widehat M^{(z_{1,1},z_2)}: \beta_1^{-1} \beta_{p+2} \gamma_1 \gamma_{p+2},\; \widehat M^{(z_{1,2},z_2)}: \beta_1^{-1} \beta_{p+2}^{-1} \gamma_1 \gamma_{p+2}^{-1},\; \{ \widehat M^{(z_j,z_{j+1})}: \{ \beta_{j}^{-1} \gamma_j\} \}_{j=2}^{p}, \nonumber\\
 & & \ \,  \widehat M^{(z_{p+1},z_{p+2,1})}: \beta_{p+1}^{-1} \beta_{p+3} \gamma_{p+1} \gamma_{p+3},\; \widehat M^{(z_{p+1},z_{p+2,2})}: \beta_{p+1}^{-1} \beta_{p+3}^{-1} \gamma_{p+1} \gamma_{p+3}^{-1}\}
\,.
\ee
The two maximal punctures have different charges of the moment map operators, and therefore of a different type.\footnote{The two maximal punctures actually differ by having the opposite charges of the moment map operators except for $U(1)_{\beta_{p+2}}$ and $U(1)_{\beta_{p+3}}$, this is often referred to as two punctures differing by a sign. Two such punctures can be glued to one another after identifying oppositely $U(1)_{\beta_{p+2}}$ and $U(1)_{\beta_{p+3}}$ by gauging the diagonal subgroup of their associated symmetries ($S$-gluing).} These maximal punctures in addition break the $SO(2p+6)^2$ symmetry of the $6d$ theory to its Cartan subalgebra denoted by $U(1)^{p+3}_\beta \times U(1)^{p+3}_\gamma$. 

Gluing two maximal punctures using the so called $\Phi$-gluing is done by identifying two maximal punctures of the same type and gauging their diagonal $SU(N)^4 \times SU(2N)^p$ symmetry. In addition one needs to add four bifundamental fields, one between each of the $SU(2N)$ nodes at the edges of the quiver and their two neighboring $SU(N)$ nodes, and also add $p-1$ bifundamental fields one between each neighboring $SU(2N)$ nodes. Thus, we add $p+3$ fields $\Phi_i$, coupled through the superpotential as follows,
\be
\label{E:DPhiSP}
W=\sum_{i=1}^{p+3} \Phi_i \left(\widehat M_i^{(X)} - \widehat N_i^{(X)}\right)\,,
\ee
where $\widehat M_i^{(X)}$ and $\widehat N_i^{(X)}$ are the two moment maps of the two punctures.

We will also employ another type of gluing named $S$-gluing.\footnote{For more examples of $S$-gluing see \cite{Bah:2012dg,Hanany:2015pfa}.} This gluing is used between two punctures of different types, specifically that have moment maps with exactly opposite charges,\footnote{One can consider S-gluing between punctures of the same type, but this requires identifying the charges on the two sides of the gluing oppositely. This is only possible without breaking internal symmetries when gluing two punctures of different surfaces, for example two maximal punctures on two different trinions.} and gauging their diagonal $SU(N)^4 \times SU(2N)^p$ symmetry. In addition one needs to couple their respective moment maps with the superpotential,
\be
\label{E:DSSP}
W=\sum_{i=1}^{p+3} \widehat M_i^{(X)} \widehat N_i^{(X)}\,.
\ee

To demonstrate these gluings we write the index of a four punctured sphere with two maximal punctures and two minimal punctures built by $\Phi$-gluing the two trinions along a $\textbf{z}$ type of puncture,
\be
\label{E:MinDNFourPz}
\mathcal{I}_{\boldsymbol{v},\boldsymbol{u};\epsilon,\delta}^{(N,p)} & = & \left(\frac{\kappa^{N-1}}{N!}\right)^{4}\left(\frac{\kappa^{2N-1}}{(2N)!}\right)^{p}\prod_{I=1}^{N}\prod_{n=1}^{2}\oint\frac{dz_{1,n}^{(I)}}{2\pi iz_{1,n}^{(I)}}\oint\frac{dz_{p+2,n}^{(I)}}{2\pi iz_{p+2,n}^{(I)}}\prod_{i=1}^{2N}\prod_{a=2}^{p+1}\oint\frac{dz_{a}^{(i)}}{2\pi iz_{a}^{(i)}}\times\nonumber\\
 & & \mathcal{I}_{\boldsymbol{z},\boldsymbol{v},\epsilon}^{T(N,p)}\mathcal{I}_{\boldsymbol{z},\boldsymbol{u},\delta}^{T(N,p)}\frac{1}{\prod_{I\ne J}^{N}\prod_{n=1}^{2}\Gamma_{e}\left(z_{1,n}^{(I)}\left(z_{1,n}^{(J)}\right)^{-1}\right)\Gamma_{e}\left(z_{p+2,n}^{(I)}\left(z_{p+2,n}^{(J)}\right)^{-1}\right)}\times\nonumber\\
 & & \prod_{j=1}^{2N}\prod_{I=1}^{N}\Gamma_{e}\left(\left(pq\right)^{\frac{1}{2}}\frac{\beta_{1}}{\beta_{p+2}\gamma_{1}\gamma_{p+2}}\frac{z_{2}^{(j)}}{z_{1,1}^{(I)}}\right)\Gamma_{e}\left(\left(pq\right)^{\frac{1}{2}}\frac{\beta_{1}\beta_{p+2}\gamma_{p+2}}{\gamma_{1}}\frac{z_{2}^{(j)}}{z_{1,2}^{(I)}}\right)\times\nonumber\\
 & & \prod_{i=1}^{2N}\prod_{J=1}^{N}\Gamma_{e}\left(\left(pq\right)^{\frac{1}{2}}\frac{\beta_{p+1}}{\beta_{p+3}\gamma_{p+1}\gamma_{p+3}}\frac{z_{p+2,1}^{(J)}}{z_{p+1}^{(i)}}\right)\Gamma_{e}\left(\left(pq\right)^{\frac{1}{2}}\frac{\beta_{p+1}\beta_{p+3}\gamma_{p+3}}{\gamma_{p+1}}\frac{z_{p+2,2}^{(J)}}{z_{p+1}^{(i)}}\right)\times\nonumber\\
 & & \frac{1}{\prod_{i\ne j}^{2N}\prod_{a=2}^{p+1}\Gamma_{e}\left(z_{a}^{(i)}\left(z_{a}^{(j)}\right)^{-1}\right)}\prod_{i,j=1}^{2N}\prod_{a=2}^{p}\Gamma_{e}\left(\left(pq\right)^{\frac{1}{2}}\frac{\beta_{a}}{\gamma_{a}}\frac{z_{a+1}^{(j)}}{z_{a}^{(i)}}\right).
\ee
Demonstrating in a similar fashion the $S$-gluing we show the index of a four punctured sphere with two maximal punctures of type $\textbf{z}$ and $\textbf{u}$ and two minimal punctures built by $S$-gluing two trinions along a $\textbf{z}$ type puncture in one and a $\textbf{u}$ type puncture in the other,\footnote{Remember that one of the trinions need to be with flipped $U(1)_{\beta_{p+2}}$ and $U(1)_{\beta_{p+3}}$ charges.}
\be
\mathcal{I}_{\boldsymbol{z};\boldsymbol{u};\epsilon,\delta}^{(N,p)}&=&\left(\frac{\kappa^{N-1}}{N!}\right)^{4}\left(\frac{\kappa^{2N-1}}{(2N)!}\right)^{p}\prod_{I=1}^{N}\prod_{n=1}^{2}\oint\frac{dv_{1,n}^{(I)}}{2\pi iv_{1,n}^{(I)}}\oint\frac{dv_{p+2,n}^{(I)}}{2\pi iv_{p+2,n}^{(I)}}\prod_{i=1}^{2N}\prod_{a=2}^{p+1}\oint\frac{dv_{a}^{(i)}}{2\pi iv_{a}^{(i)}}\times\nonumber\\
 & & \frac{\mathcal{I}_{\boldsymbol{z},\boldsymbol{v},\epsilon}^{T(N,p)}\mathcal{I}_{\boldsymbol{v},\boldsymbol{u},\delta}^{T(N,p)}\left(\beta_{p+2}\to\beta_{p+2}^{-1},\beta_{p+3}\to\beta_{p+3}^{-1}\right)}{\prod_{I\ne J}^{N}\prod_{n=1}^{2}\Gamma_{e}\left(v_{1,n}^{(I)}\left(v_{1,n}^{(J)}\right)^{-1}\right)\Gamma_{e}\left(v_{p+2,n}^{(I)}\left(v_{p+2,n}^{(J)}\right)^{-1}\right)}\times\nonumber\\
 & & \frac{1}{\prod_{i\ne j}^{2N}\prod_{a=2}^{p+1}\Gamma_{e}\left(v_{a}^{(i)}\left(v_{a}^{(j)}\right)^{-1}\right)}
\ee

\subsection{Checks}
The new trinion can be validated by several checks we can preform. First it would have been nice to associate the new minimal puncture to a known maximal puncture, as a partial closure of this maximal puncture by giving vev to operators charged under it. Unfortunately we could not find such a maximal puncture and it seems that the known maximal punctures of this class with symmetry $SU(N)^4 \times SU(2N)^p$ are not associated with the new minimal puncture. If such a maximal puncture exists we might expect it to be a generalization of the $USp(2p)$ puncture in the case of $N=1$ as was found in \cite{Razamat:2019ukg}. 

\begin{figure}[t]
	\centering
  	\includegraphics[scale=0.3]{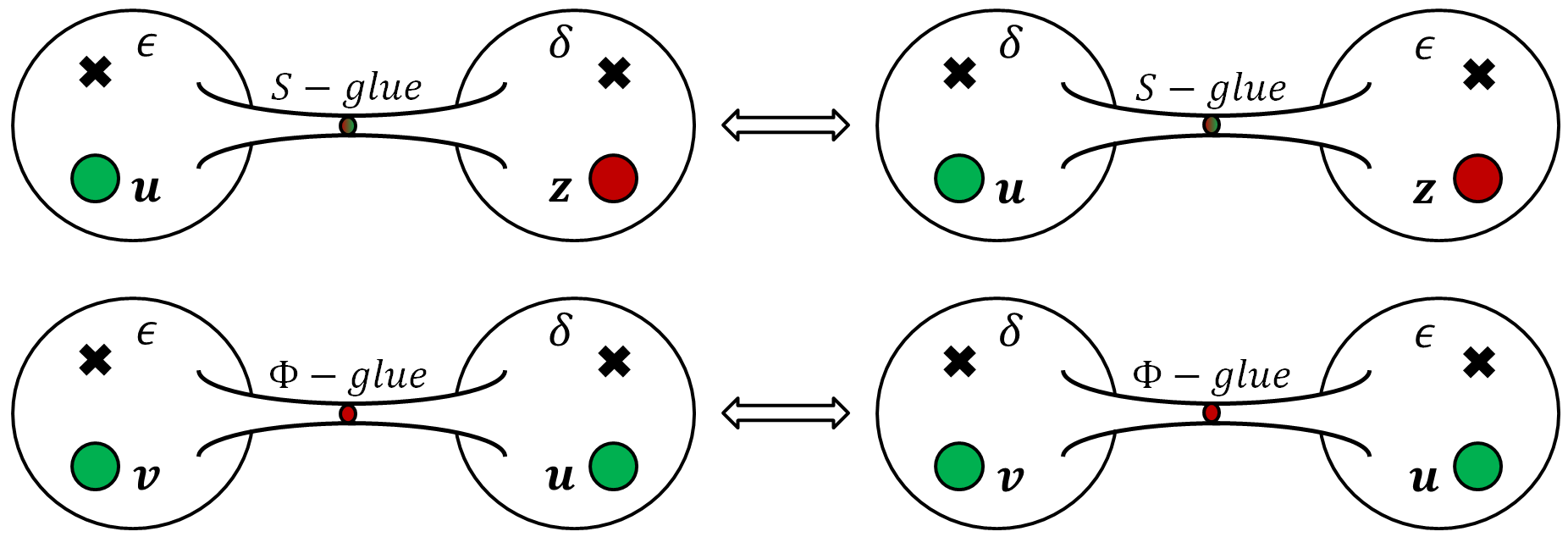}
    \caption{Different duality frames for a four punctured sphere. The fact that the left and right frames are the same implies for example that the index has to be invariant under exchange of the two $U(1)$ fugacities $\epsilon$ and $\delta$. }
    \label{F:NonMinDDuality}
\end{figure}

Nevertheless, there are several checks we can preform. One non-trivial check we can preform on the conjectured trinion is to show that models with more than three punctures satisfy duality properties. One such property is showing that the index is invariant under the exchange of two punctures of the same type, see Figure  \ref{F:NonMinDDuality}. We have proved this property using a series of Seiberg and S-dualities for the case of $p=1$ in Appendix \ref{A:MinPuncDuality}. In addition we have verified this property by using an expansion in fugacities for $p>1$.\footnote{As for the $p=1$ case, we expect that for $p>1$ the relevant identity satisfied by the index can be deduced from sequences of Seiberg and S-dualities.}

Another check we can preform is to close the new minimal puncture by giving a vev to operators charged under it in a similar manner to closing punctures in other previously studied setups \cite{Gaiotto:2015usa}. By examining such analogous cases we expect the operators to be the unflipped baryonic operators charged under the new minimal puncture symmetry.
We expect after closing the minimal puncture and adding some singlet flip fields in the process that the resulting theory will be a known flux tube theory \cite{Kim:2018bpg}. The flux associated to such tubes should be predicted by the veved operator charges.

Now, we consider giving a vev to the above baryonic operators, there are $2p+6$ options all with R-charge $N$. These operators charges are $\epsilon^{2N}\beta_i^{2N}$ for $i=1,...,p+1$, $\epsilon^{-2N}\beta_{p+2}^{-2N}$, $\epsilon^{-2N}\beta_{p+3}^{-2N}$, $\epsilon^{-2N}\gamma_j^{-2N}$ for $j=1,...,p+1$, $\epsilon^{-2N}\gamma_{p+2}^{2N}$ and $\epsilon^{-2N}\gamma_{p+3}^{2N}$. Closing the puncture by giving a vev to one of these operators shifts the flux of the theory by one quanta opposite to the internal symmetries charges of the veved operator. For instance, giving a vev to an operator with charges $\epsilon^{2N}\beta_i^{\pm 2N}$ shifts the flux of the trinion by $\mp 1$ for $U(1)_{\beta_i}$. As stated above, we also need to add some singlet flip fields. These will be determined such that the resulting theory 't Hooft anomalies will match the ones predicted from six dimensions. We will only specify the flippings required when setting vevs for the operators $\epsilon^{2N}\beta_i^{2N}$ for $i=1,...,p+1$ and  $\epsilon^{-2N}\gamma_j^{-2N}$ for $j=2,...,p$, as the others are a bit different and are not required for this check. We find that one needs to couple flip fields to all the baryonic operators $\epsilon^{2N}\beta_i^{2N}$ with $i=1,...,p+1$ and  $\epsilon^{-2N}\gamma_j^{-2N}$ with $j=2,...,p$ except the veved one, and also flip the operator of $2-N$ R-charge, same $\epsilon$ charge and opposite $\beta_i$ or $\gamma_j$ charges as the veved operator. In addition, one need to flip the flipping fields $F_{i4}$ and $\widetilde{F}_{ij}$ with $i,j=1,2,3$.\footnote{Flipping a flip field simply amounts to giving it a mass.} These flipping fields are enough to match the anomalies predicted form $6d$.

\begin{figure}[t]
	\centering
  	\includegraphics[scale=0.32]{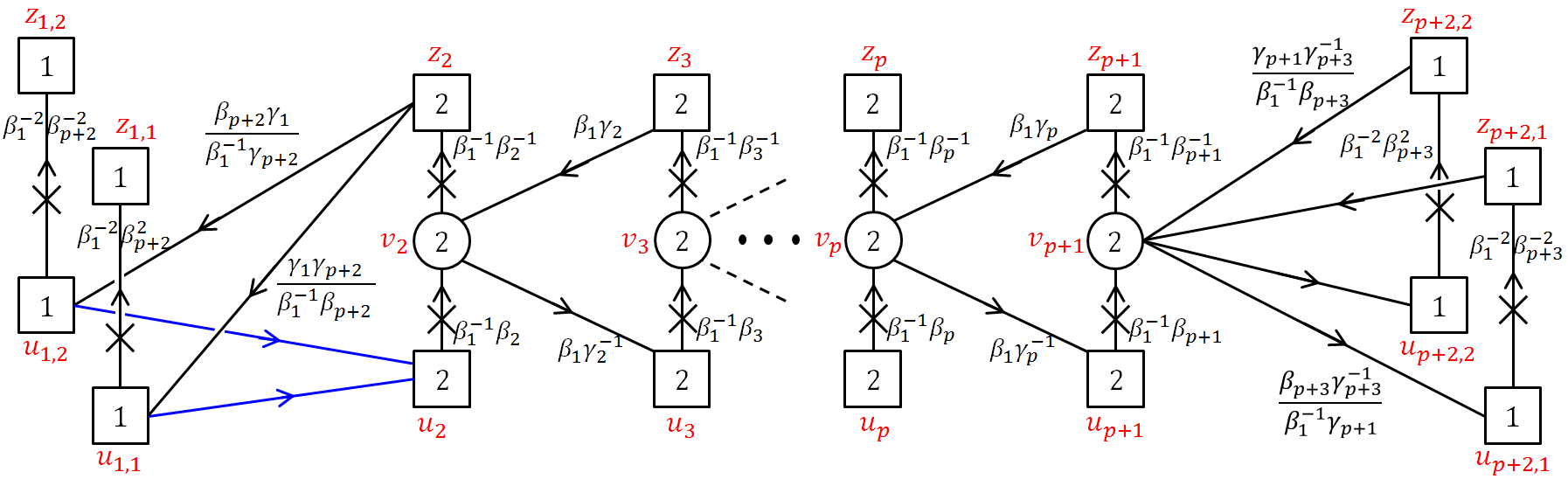}
    \caption{A quiver diagram describing the IR theory one finds after closing the minimal puncture of the trinion by giving a vev to the operator $\widetilde M_1^{2N}$. The squares and circles denote $SU(nN)$ global and gauge symmetries, respectively, where $n$ is the number inside. The fields denoted by the vertical lines have a vanishing R-charge, the flip fields have R-charge $2$ and the rest of the fields have R-charge $1$. $\beta_i$ and $\gamma_j$ are related to the internal symmetries that arise from $6d$. In red we write the symmetries associated fugacities, and in black we write the charges of each field. The $X$'s denote flip fields. We emphasize that the six additional flip fields charged under $c$ and $\widetilde c$ were removed. As always, each closed loop of fields has a  superpotential term turned on for it.}
    \label{F:NonMinDNTube}
\end{figure}

To give a concrete example, we choose to close the minimal puncture of the trinion by giving a vev to the baryonic operator $\widetilde{M}_1^{2N}$ with charges $\epsilon^{2N} \beta_1^{2N}$. This generates an RG flow resulting in the IR theory described in the quiver diagram of Figure \ref{F:NonMinDNTube}.
By construction the remaining theory has two maximal punctures. This flux tube has a flux of $-1$ for $U(1)_{\beta_1}$ and a vanishing flux for the rest of the $U(1)$'s.\footnote{The flux conventions used here are of opposite sign from the ones used in \cite{Kim:2018lfo}, as these are more natural in the derivation of the anomaly polynomial from $6d$ as shown in Appendix \ref{A:Anomalies}.}

Next, we $\Phi$-glue two such tubes to generate a flux torus, and check these are the expected anomalies from $6d$. We find the following anomalies,
\be
&& Tr\left(U(1)_{\beta_{1}}\right)=8N(p+2)\,,\qquad Tr\left(U(1)_{\beta_{1}}^{3}\right)=16N^{2}\left(2N\left(p+2\right)-3\right)\,,\nonumber\\
&& Tr\left(U(1)_{R}^{2}U(1)_{\beta_{1}}\right)=-8N\left(2N\left(p+1\right)-p-2\right)\,,\nonumber\\
&& Tr\left(U(1)_{\beta_{1}}U(1)_{\beta_{i\ne1}}^{2}\right)=16N^{2}\left(2N-1\right)\,,\qquad Tr\left(U(1)_{\beta_{1}}U(1)_{\gamma_{i}}^{2}\right)=16N^{2}\,,
\ee
where the rest of the anomalies vanish. These anomalies exactly match the expectations form $6d$ given in Appendix \ref{A:Anomalies} for a torus of flux $-2$ for $U(1)_{\beta_1}$ and zero for the rest of the $U(1)$'s.

Finally, one can check that the above conjectured trinion reduces to the known trinion of the $D_{p+3}$ minimal conformal matter \cite{Razamat:2019ukg} when we set $N=1$. In addition to setting $N=1$ we will also change to the matching notation where we take $\epsilon \to \epsilon^{1/2}$, $\beta_i = t a_i$ for $i=1,...,p+1$ with $\prod_{i=1}^{p+1} a_i = 1$, $\gamma_j = s_{j-1}$ for $j=2,...,p$ and also
\be
c_{n} & = & \left(\beta_{p+2}\gamma_{1}\gamma_{p+2},\beta_{p+2}^{-1}\gamma_{1}\gamma_{p+2}^{-1},\beta_{p+2}\gamma_{1}^{-1}\gamma_{p+2}^{-1},\beta_{p+2}^{-1}\gamma_{1}^{-1}\gamma_{p+2}\right)\,,\nonumber\\
\widetilde{c}_{n} & = & \left(\beta_{p+3}\gamma_{p+1}\gamma_{p+3},\beta_{p+3}^{-1}\gamma_{p+1}\gamma_{p+3}^{-1},\beta_{p+3}\gamma_{p+1}^{-1}\gamma_{p+3}^{-1},\beta_{p+3}^{-1}\gamma_{p+1}^{-1}\gamma_{p+3}\right)\,,
\ee
with $\prod_{n=1}^{4} c_i = \prod_{n=1}^{4} \widetilde{c}_i = 1$. In addition we switch $v_{i}\to y_{i},\,u_{i}\to v_{i-1}$, and $z_{i}\to z_{i-1}$. Using the above notations the trinion index in \eqref{E:DTrinion} reduces to
\be
\mathcal{I}_{\boldsymbol{z},\boldsymbol{v},\epsilon}^{T(N=1,p)} & = & \kappa^{p+1}\prod_{i=1}^{p+1}\oint\frac{dy_{i}}{4\pi iy_{i}}\frac{\prod_{n=1}^{4}\Gamma_{e}\left(\left(pq\right)^{1/4}\epsilon^{-1/2}y_{1}^{\pm1}c_{n}^{-1}\right)\Gamma_{e}\left(\left(pq\right)^{1/4}\epsilon^{-1/2}y_{p+1}^{\pm1}\widetilde{c}_{n}\right)}{\prod_{i=1}^{p+1}\Gamma_{e}\left(y_{i}^{\pm2}\right)}\times\nonumber\\
 & & \prod_{n=1}^{2}\Gamma_{e}\left(\sqrt{pq}t^{-1}a_{1}^{-1}z_{1}^{\pm1}c_{n}\right)\prod_{n=3}^{4}\Gamma_{e}\left(\sqrt{pq}ta_{1}v_{1}^{\pm1}c_{n}\right)\times\nonumber\\
 & & \prod_{n=1}^{3}\Gamma_{e}\left(\sqrt{pq}\epsilon c_{n}c_{4}\right)\Gamma_{e}\left(\sqrt{pq}\epsilon\widetilde{c}_{n}\widetilde{c}_{4}\right)\prod_{j=1}^{p-1}\Gamma_{e}\left(\sqrt{pq}\epsilon s_{j}^{-2}\right)\prod_{i=1}^{p+1}\Gamma_{e}\left(\sqrt{pq}\epsilon^{-1}t^{2}a_{i}^{2}\right)\times\nonumber\\
 & & \Gamma_{e}\left(\left(pq\right)^{1/4}\epsilon^{1/2}ta_{1}y_{1}^{\pm1}z_{1}^{\pm1}\right)\Gamma_{e}\left(\left(pq\right)^{1/4}\epsilon^{1/2}t^{-1}a_{1}^{-1}y_{1}^{\pm1}v_{1}^{\pm1}\right)\Gamma_{e}\left(\sqrt{pq}\epsilon^{-1}v_{1}^{\pm1}z_{1}^{\pm1}\right)\times\nonumber\\
 & & \prod_{i=1}^{p}\Gamma_{e}\left(\left(pq\right)^{1/4}\epsilon^{1/2}t^{-1}a_{i+1}^{-1}z_{i}^{\pm1}y_{i+1}^{\pm1}\right)\Gamma_{e}\left(\left(pq\right)^{1/4}\epsilon^{1/2}ta_{i+1}v_{i}^{\pm1}y_{i+1}^{\pm1}\right)\times\nonumber\\
 & & \prod_{j=1}^{p-1}\Gamma_{e}\left(\left(pq\right)^{1/4}\epsilon^{-1/2}s_{j}y_{j+1}^{\pm1}z_{j+1}^{\pm1}\right)\Gamma_{e}\left(\left(pq\right)^{1/4}\epsilon^{-1/2}s_{j}^{-1}y_{j+1}^{\pm1}v_{j+1}^{\pm1}\right)\,.
\ee

Finally, we need to use Seiberg duality on the $SU(2)_{y_1}$ gauge node to get the index to look the same as in \cite{Razamat:2019ukg}.\footnote{The duality frame selected only exists for the $N=1$ case, as the gauge symmetry in this case has only pseudo-real representations.} For this duality we choose the fields charged under $c_n$ as the fundamental and the rest as the antifundamental, and we find
\be
\mathcal{I}_{\boldsymbol{z},\boldsymbol{v},\epsilon}^{T(N=1,p)} & = & \kappa^{p+1}\prod_{a=1}^{p+1}\oint\frac{dy_{a}}{4\pi iy_{a}}\frac{\prod_{n=1}^{4}\Gamma_{e}\left(\left(pq\right)^{1/4}\epsilon^{-1/2}y_{1}^{\pm1}c_{n}\right)\Gamma_{e}\left(\left(pq\right)^{1/4}\epsilon^{-1/2}y_{p+1}^{\pm1}\widetilde{c}_{n}\right)}{\prod_{a=1}^{p+1}\Gamma_{e}\left(y_{a}^{\pm2}\right)}\times\nonumber\\
 & & \prod_{n=3}^{4}\Gamma_{e}\left(\sqrt{pq}t^{-1}a_{1}^{-1}z_{1}^{\pm1}c_{n}^{-1}\right)\prod_{n=1}^{2}\Gamma_{e}\left(\sqrt{pq}ta_{1}v_{1}^{\pm1}c_{n}^{-1}\right)\times\nonumber\\
 & & \prod_{n=1}^{3}\Gamma_{e}\left(\sqrt{pq}\epsilon c_{n}c_{4}\right)\Gamma_{e}\left(\sqrt{pq}\epsilon\widetilde{c}_{n}\widetilde{c}_{4}\right)\prod_{j=1}^{p-1}\Gamma_{e}\left(\sqrt{pq}\epsilon s_{j}^{-2}\right)\prod_{i=1}^{p+1}\Gamma_{e}\left(\sqrt{pq}\epsilon^{-1}t^{2}a_{i}^{2}\right)\times\nonumber\\
 & & \Gamma_{e}\left(\left(pq\right)^{1/4}\epsilon^{1/2}t^{-1}a_{1}^{-1}y_{1}^{\pm1}z_{1}^{\pm1}\right)\Gamma_{e}\left(\left(pq\right)^{1/4}\epsilon^{1/2}ta_{1}y_{1}^{\pm1}v_{1}^{\pm1}\right)\Gamma_{e}\left(\sqrt{pq}\epsilon^{-1}v_{1}^{\pm1}z_{1}^{\pm1}\right)\times\nonumber\\
 & & \prod_{i=1}^{p}\Gamma_{e}\left(\left(pq\right)^{1/4}\epsilon^{1/2}t^{-1}a_{i+1}^{-1}z_{i}^{\pm1}y_{i+1}^{\pm1}\right)\Gamma_{e}\left(\left(pq\right)^{1/4}\epsilon^{1/2}ta_{i+1}v_{i}^{\pm1}y_{i+1}^{\pm1}\right)\times\nonumber\\
 & & \prod_{j=1}^{p-1}\Gamma_{e}\left(\left(pq\right)^{1/4}\epsilon^{-1/2}s_{j}y_{j+1}^{\pm1}z_{j+1}^{\pm1}\right)\Gamma_{e}\left(\left(pq\right)^{1/4}\epsilon^{-1/2}s_{j}^{-1}y_{j+1}^{\pm1}v_{j+1}^{\pm1}\right)\,.
\ee
The resulting trinion is very close to the one found in \cite{Razamat:2019ukg} of two maximal punctures of symmetry $SU(2)^p$ and one minimal puncture of symmetry $SU(2)$ only seen in the IR. The only difference are the fields appearing in the second line of the formula, who simply flip some operators. These can be seen as a different choice of boundary conditions for the maximal punctures. This concludes the final check for the new trinion, as it reproduces the known trinion of $N=1$.\footnote{Notice that in \cite{Razamat:2019ukg} $p$ is exchanged with $N$.}

\section{The trinion derivation from RG flows}\label{S:Derivation}
In this section we will derive the trinion with two $SU(N)^4 \times SU(2N)^p$ maximal punctures and one $U(1)$ minimal puncture for the  $D_{p+3}$ non-minimal conformal matter. We will first summarize the understandings of \cite{Razamat:2019mdt}, as the derivation will be heavily dependent on them. Then, we will use these understandings to derive the trinion by initiating a flow from $D_{p+4}$ non-minimal conformal matter compactified on a torus with flux to $D_{p+3}$ non-minimal conformal matter compactified on a torus with flux and extra minimal punctures. The resulting model will be identified as several flux tubes glued to the aforementioned trinion. the derivation will be shown in full detail only for $p=1$, to avoid unnecessary overclouding of the main idea. For higher values of $p$ the derivation will follow exactly the same steps; thus, can be easily generalized.

\subsection{From $6d$ flows to $4d$ flows}

\begin{figure}[t]
	\centering
  	\includegraphics[scale=0.31]{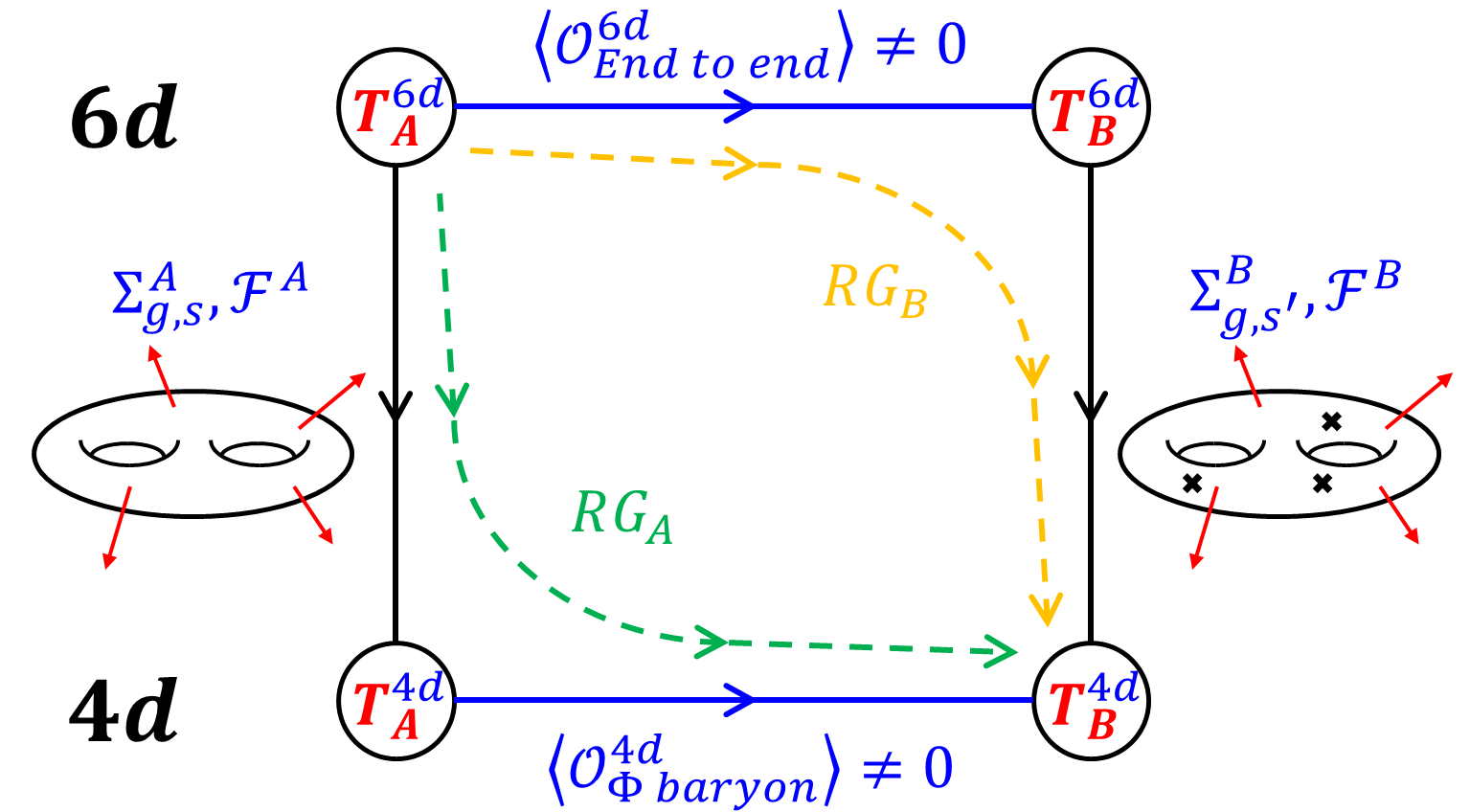}
    \caption{A diagram representing the two different orders of RG flows that were considered in \cite{Razamat:2019mdt}. Flow $RG_A$ describes the $6d\to 4d \to 4d$ path where a compactification of a $6d$ SCFT to an effective $4d$ theory is followed by a vev to an operator in $4d$. Flow $RG_B$ describes the $6d\to 6d \to 4d$ path where one first triggers the vev to a $6d$ operator and then compactify the model to $4d$.}
    \label{F:6dRGto4dRG}
\end{figure}

\begin{figure}[t]
	\centering
  	\includegraphics[scale=0.31]{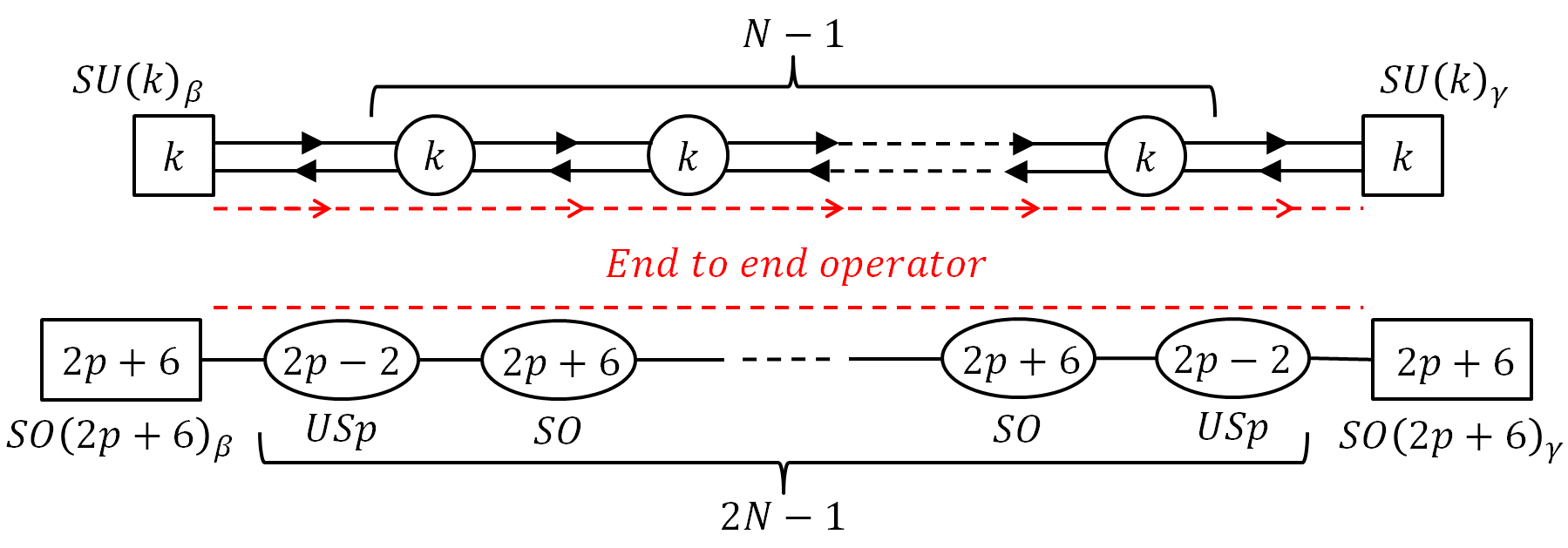}
    \caption{Quiver diagrams of the tensor branch theories of the $6d$ $(1,0)$ $\mathcal{T}\left(SU(k),N\right)$ (above) and $\mathcal{T}\left(SO(2p+6),N\right)$ (below) SCFTs. The arrows represent half hyper-multiplets in the bifundamental representation of $SU(k)\times SU(k)$, while lines represent half hyper-multiplets in the $(\textbf{2N+6},\textbf{2N-2})$ representation of $SO(2N+6)\times USp(2N-2)$. In the lower quiver there are $2N-1$ gauge nodes interchanging between $SO(2p+6)$ and $USp(2p-2)$. The dashed red line represents the "end to end" operators of each SCFT.}
    \label{F:End2EndOp}
\end{figure}

In \cite{Razamat:2019mdt} the authors consider $6d$ $(1,0)$ SCFTs denoted by $\mathcal{T}\left(SU(k),N\right)$. These SCFTs can be described as the low energy limit of a stack of M5-branes probing a $\mathbb{Z}_k$ singularity. Two types of flows are considered for these SCFTs. The first, is a geometric flow generated by compactifying the theory on a Riemann surface with fluxes to $4d$. This type of flow results in a class of theories denoted as class $\mathcal{S}_k$ \cite{Gaiotto:2015usa}. The second type of flow is generated by giving a vev to a $6d$ operator that winds between the $6d$ tensor branch quiver two ends, see Figure \ref{F:End2EndOp}. This $6d$ operator was referred to as the ``end to end'' operator, and it is charged in the fundamental representation of one of the flavor $SU(k)$ and the antifundamental of the other $SU(k)$. The flow triggered by giving a vev to such an operator reduces $k$ resulting in the 6d SCFT denoted by $\mathcal{T}(SU(k-1),N)$. These two flows were considered in two different orders. In the first denoted by $6d\to 6d \to 4d$, we first trigger the vev in $6d$ and then compactify the theory ending in a $4d$ model. In the second denoted by $6d\to 4d \to 4d$, we first compactify to $4d$ and then trigger a vev to a $4d$ operator ending with the same $4d$ model as before, see Figure \ref{F:6dRGto4dRG}. In \cite{Razamat:2019mdt} a nontrivial mapping between the two flow orders was found, which is the foundation for the derivation of our new models.

We can think of the two deformations leading to the RG-flows as not being strictly ordered in one way or another. Instead each deformation has an energy scale related to the scale of the vev and the geometry size, and these can be deformed smoothly from $6d\to 6d \to 4d$ to $6d\to 4d \to 4d$ by changing these energy scales. Thus, both deformations are ``turned on'' simultaneously and need to be considered together, and this will be the approach from here on out. Due to this reason one can expect that the two strictly ordered flows can be mapped to one another in a manner that leads to the same $4d$ model outcome. 

In order to map these flows, we first need to find the $4d$ operator arising from the $6d$ ``end to end'' operator under the compactification. Assuming the flux is general we expect the $SU(k)_\beta\times SU(k)_\gamma$ global symmetry of the $6d$ SCFT to be generally broken to its Cartan symmetry. Thus, we still expect the required $4d$ operator to be charged under $U(1)_{\beta_i}$ and $U(1)_{\gamma_j}$ with opposite charges, and in addition have the same charges as the $6d$ ``end to end'' operator under the rest of the symmetries.

Next, we need to match the Riemann surfaces and fluxes of the two flows. In \cite{Razamat:2019mdt} it was argued that if the flux is being turned on for symmetries that the $6d$ ``end to end'' operator is charged under one cannot turn on a constant vev for this operator, and the vev needs to vary along the compactified directions. Using such a space dependent vev it was argued with brane constructions and field theory techniques that the vev spatial profile can localize on points of the compactification surface, and can be interpreted as additional punctures. These punctures were each associated with a $U(1)$ symmetry. The implications for $4d$ class $\mathcal{S}_k$ models are that by triggering a vev to a $4d$ operator matching the $6d$ ``end to end'' operator we can flow to a theory of class $\mathcal{S}_{k-1}$ described by a new Riemann surface with extra minimal punctures compared to the original surface. The number of extra punctures as well as the new flux will be related to the original theory flux, and can be deduced in various ways including anomalies matching to the ones predicted from $6d$.  

Generating extra punctures by a vev driven RG-flow has been considered for the $4d$ compactifications of $\mathcal{T}\left(SU(k),N\right)$ \cite{Razamat:2019mdt} and $\mathcal{T}\left(SO(2N+6),1\right)$ \cite{Razamat:2019ukg}. The reasoning behind these processes can be similarly followed for the $6d$ $(1,0)$ SCFTs denoted by $\mathcal{T}\left(SO(2p+6),N\right)$. These SCFTs can be described by a stack of $N$ M5-branes probing a $D_{p+3}$ singularity. In the next subsection we will consider these models and their compactifications to $4d$ generalizing the derivation for $N=1$ that was done in \cite{Razamat:2019ukg}.

\subsection{Generating extra punctures in $D_{p+3}$ conformal matter compactifications using RG flows}
\begin{figure}[t]
	\centering
  	\includegraphics[scale=0.31]{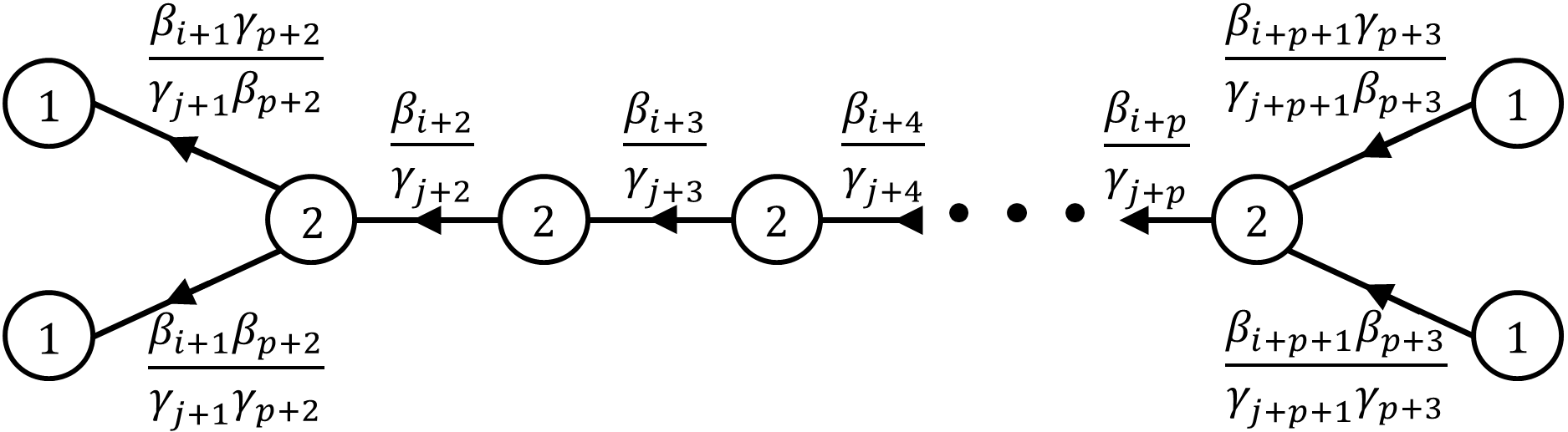}
    \caption{The fields added in $\Phi$-gluing. The circles denote gauge nodes of gauge symmetry $SU(nN)$ where $n$ is the number inside the circle. The $\beta_i$ and $\gamma_j$ fugacities are defined cyclically such that $\beta_{p+2},\beta_{p+3}$ and $\gamma_{p+2},\gamma_{p+3}$ can only get exchanged with their multiplicative inverse, while the rest are defined as $\beta_i\equiv \beta_{(i \mod (p+1))+1}$ and $\gamma_i\equiv \gamma_{(i \mod (p+1))+1}$. The baryonic operators with charges $\beta_i^{2N} \gamma_j^{-2N}$ introduced in the gluing have matching charges to the $6d$ ``end to end'' operators and are the ones we give vacuum expectation value to. Only some of the operators exist in every model, since the spectrum depends on the fluxes and also puncture properties in case there are ones. Therefore, the flow is expected to depend non trivially on the fluxes. All fields added in the gluing have $+1$ R-charge. This R-charge is the one naturally inherited from $6d$ and not necessarily the conformal one. Notice that the baryonic operators in the edges need to be built from both of the edge fields.}
    \label{F:DPhiGluingFields}
\end{figure}

Here we will apply the understandings of \cite{Razamat:2019mdt} as summarized above to $\mathcal{T}\left(SO(2p+6),N\right)$. This will be done in analogous manner to the derivation of \cite{Razamat:2019ukg}. The $6d$ ``end to end'' operators for $\mathcal{T}\left(SO(2p+6),N\right)$ are the ones that as expected wind from one end of the $6d$ tensor branch quiver to the other, as shown in Figure \ref{F:End2EndOp}. These $6d$ operators have $4d$ counterparts with the same charges under the internal symmetries, and just as in the minimal case and the $A$-type case, are baryonic operators built from the $\Phi$ fields added when $\Phi$-gluing (see Figure \ref{F:DPhiGluingFields} for a quiver illustration of the added fields).

The derivation is similar to the one in \cite{Razamat:2019ukg}, where in the first part one needs to identify the internal symmetries of class $\mathcal{S}_{D_{p+2}}$ from the ones of class $S_{D_{p+3}}$. This identification can be done by starting with two flux tubes $\Phi$-glued to one another in class $S_{D_{p+3}}$ and initiating the aforementioned flow by giving a vev to one of the baryonic operators built from one of the $\Phi$ fields added in the $\Phi$-gluing. This flow is expected to end in a similar model only for class $\mathcal{S}_{D_{p+2}}$ as seen before in both the $A$-type flows and minimal $D$-type. For the general case of a flow generated by giving a vev to a baryonic operator of charges $\beta_i^{2N} \gamma_j^{-2N}$, one finds the identification of the internal symmetries is $\beta_{\ell<i}^{new}=\beta_\ell,\, \beta_{\ell\ge i}^{new}=\beta_{\ell+1}$ and $\gamma_{\ell<j}^{new}=\gamma_\ell,\, \gamma_{\ell\ge j}^{new}=\gamma_{\ell+1}$. 

\begin{figure}[t]
	\centering
  	\includegraphics[scale=0.31]{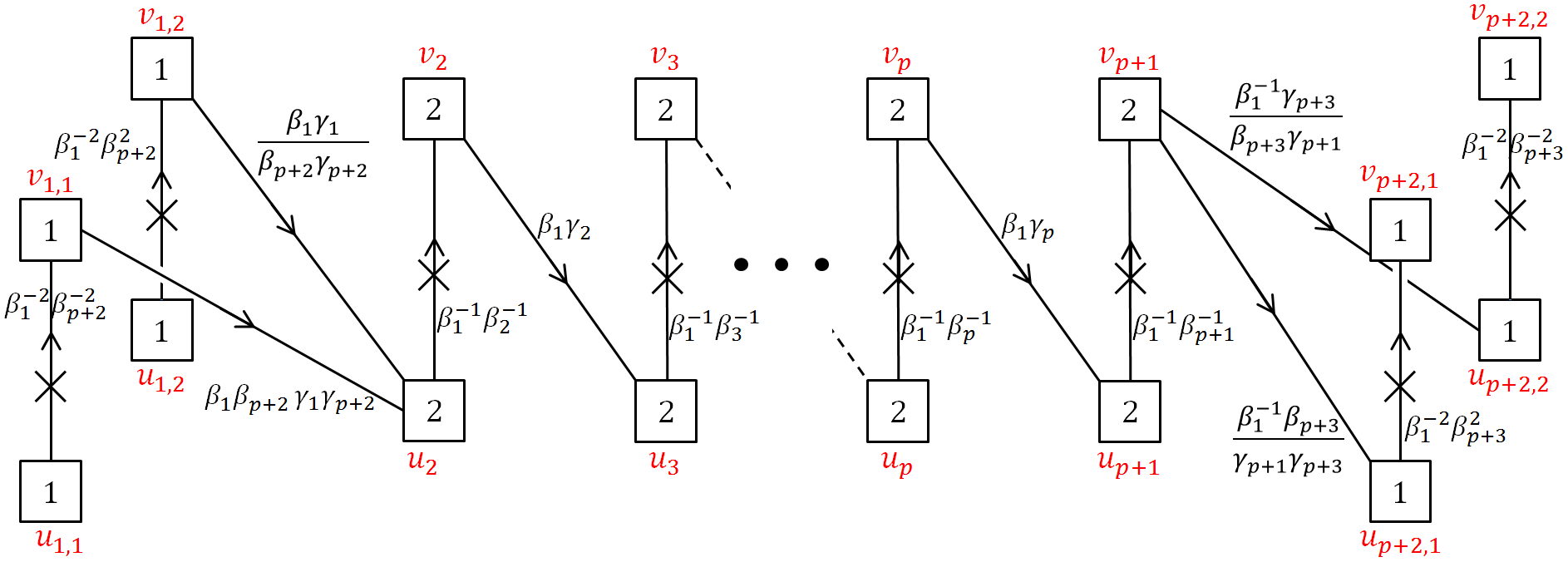}
    \caption{A flux tube quiver with fluxes $\mathcal{F}_{\beta_1}=\frac{p+4}{2p+4}$, $\mathcal{F}_{\beta_i}=\frac{1}{2p+4}$ for $i=2,...,p+1$ and the rest vanishing. The squares denote flavor symmetry nodes of symmetry $SU(nN)$ where $n$ is the number inside the square. We give the fields denoted by diagonal arrows R-charge $1$, the fields denoted by perpendicular arrows R-charge $0$ and the flip fields denoted by crosses R-charge $2$. This R-charge is the one naturally inherited from $6d$ and not necessarily the conformal one.}
    \label{F:FluxTube}
\end{figure}

Next, we will employ the same flow to a torus build from fundamental flux tubes $\Phi$-glued together. We will use the fundamental flux tube of fluxes $\mathcal{F}_{\beta_1}=\frac{p+4}{2p+4}$ and $\mathcal{F}_{\beta_i}=\frac{1}{2p+4}$ for $i=2,...,p+1$ with the rest vanishing. The quiver diagram of this flux tube appears in Figure \ref{F:FluxTube}. We glue such a fundamental flux tube to the next one where we shift in the next tube in the following manner \footnote{Note that the tube flux is shifted in an equivalent manner.}
\be
\left(\beta_1,\beta_2,...,\beta_{p},\beta_{p+1},\beta_{p+2},\beta_{p+3}\right) \to \left(\beta_2,\beta_3,...,\beta_{p+1},\beta_1,\beta_{p+2}^{-1},\beta_{p+3}^{-1}\right).
\ee
In total we glue $p+3$ fundamental tubes in such a manner to a torus if $p$ is odd, and $2p+6$ tubes if $p$ is even to preserve all internal symmetries \cite{Kim:2018lfo}.

Here we will give an explicit example flowing from $p=2$ to $p=1$ for simplicity. Thus, we consider $p=2$ six fundamental tubes $\Phi$-glued to a torus. The torus flux is $\mathcal{F}_{\beta_i}=2$ for $i=1,2,3$, and a vanishing flux for all the other internal symmetries, and its superconformal index is
\be
\mathcal{I}_{g=1}^{N,p=2} & = & \left[\left(\frac{\kappa^{N-1}}{N!}\right)^{4}\left(\frac{\kappa^{2N-1}}{(2N)!}\right)^{2}\prod_{I=1}^{N}\oint\frac{du_{1,1,1}^{(I)}}{2\pi iu_{1,1,1}^{(I)}}\oint\frac{du_{1,1,2}^{(I)}}{2\pi iu_{1,1,2}^{(I)}}\oint\frac{du_{1,4,1}^{(I)}}{2\pi iu_{1,4,1}^{(I)}}\oint\frac{du_{1,4,2}^{(I)}}{2\pi iu_{1,4,2}^{(I)}}\right.\nonumber\\
 & & \prod_{i=1}^{2N}\oint\frac{du_{1,2}^{(i)}}{2\pi iu_{1,2}^{(i)}}\oint\frac{du_{1,3}^{(i)}}{2\pi iu_{1,3}^{(i)}}\prod_{i,j=1}^{2N}\Gamma_{e}\left(\sqrt{pq}\beta_{2}\gamma_{2}^{-1}\left(u_{1,2}^{(i)}\right)^{-1}u_{1,3}^{(j)}\right)\nonumber\\
 & & \frac{\prod_{I,j}\Gamma_{e}\left(\sqrt{pq}\beta_{1}\beta_{4}\gamma_{1}^{-1}\gamma_{4}^{-1}\left(u_{1,1,1}^{(I)}\right)^{-1}u_{1,2}^{(j)}\right)\Gamma_{e}\left(\sqrt{pq}\beta_{1}\beta_{4}^{-1}\gamma_{1}^{-1}\gamma_{4}\left(u_{1,1,2}^{(I)}\right)^{-1}u_{1,2}^{(j)}\right)}{\prod_{i\ne j,I\ne J}\Gamma_{e}\left(u_{1,1,1}^{(I)}\left(u_{1,1,1}^{(J)}\right)^{-1}\right)\Gamma_{e}\left(u_{1,1,2}^{(I)}\left(u_{1,1,2}^{(J)}\right)^{-1}\right)\Gamma_{e}\left(u_{1,2}^{(i)}\left(u_{1,2}^{(j)}\right)^{-1}\right)}\nonumber\\
 & & \frac{\prod_{i,J}\Gamma_{e}\left(\sqrt{pq}\beta_{3}\beta_{5}\gamma_{3}^{-1}\gamma_{5}^{-1}\left(u_{1,3}^{(i)}\right)^{-1}u_{1,4,1}^{(J)}\right)\Gamma_{e}\left(\sqrt{pq}\beta_{3}\beta_{5}^{-1}\gamma_{3}^{-1}\gamma_{5}\left(u_{1,3}^{(i)}\right)^{-1}u_{1,4,2}^{(J)}\right)}{\prod_{i\ne j,I\ne J}\Gamma_{e}\left(u_{1,3}^{(i)}\left(u_{1,3}^{(j)}\right)^{-1}\right)\Gamma_{e}\left(u_{1,4,1}^{(I)}\left(u_{1,4,1}^{(J)}\right)^{-1}\right)\Gamma_{e}\left(u_{1,4,2}^{(I)}\left(u_{1,4,2}^{(J)}\right)^{-1}\right)}\nonumber\\
 & & \times\left(u_{1}\to u_{2},\beta_{1}\to\beta_{2},\beta_{2}\to\beta_{3},\beta_{3}\to\beta_{1},\beta_{4}\to\beta_{4}^{-1},\beta_{5}\to\beta_{5}^{-1}\right)\nonumber\\
 & & \left.\times\left(u_{1}\to u_{3},\beta_{1}\to\beta_{3},\beta_{2}\to\beta_{1},\beta_{3}\to\beta_{2}\right)\right]\nonumber\\
 & & \left[\Gamma_{e}\left(pq\beta_{1}^{2N}\beta_{4}^{\pm2N}\right)\Gamma_{e}\left(pq\beta_{1}^{2N}\beta_{2}^{2N}\right)\Gamma_{e}\left(pq\beta_{1}^{2N}\beta_{3}^{2N}\right)\Gamma_{e}\left(pq\beta_{1}^{2N}\beta_{5}^{\pm2N}\right)\right.\nonumber\\
 & & \prod_{j,I,J}\Gamma_{e}\left(\beta_{1}^{-2}\beta_{4}^{-2}u_{1,1,1}^{(I)}\left(u_{2,1,1}^{(J)}\right)^{-1}\right)\Gamma_{e}\left(\sqrt{pq}\beta_{1}\beta_{4}\gamma_{1}\gamma_{4}u_{2,1,1}^{(I)}\left(u_{1,2}^{(j)}\right)^{-1}\right)\nonumber\\
 & & \prod_{j,I,J}\Gamma_{e}\left(\beta_{1}^{-2}\beta_{4}^{2}u_{1,1,2}^{(I)}\left(u_{2,1,2}^{(J)}\right)^{-1}\right)\Gamma_{e}\left(\sqrt{pq}\beta_{1}\beta_{4}^{-1}\gamma_{1}\gamma_{4}^{-1}u_{2,1,2}^{(I)}\left(u_{1,2}^{(j)}\right)^{-1}\right)\nonumber\\
 & & \prod_{i,j=1}^{2N}\Gamma_{e}\left(\beta_{1}^{-1}\beta_{2}^{-1}\frac{u_{1,2}^{(i)}}{u_{2,2}^{(j)}}\right)\Gamma_{e}\left(\sqrt{pq}\beta_{1}\gamma_{2}\frac{u_{2,2}^{(i)}}{u_{1,3}^{(j)}}\right)\Gamma_{e}\left(\beta_{1}^{-1}\beta_{3}^{-1}\frac{u_{1,3}^{(i)}}{u_{2,3}^{(j)}}\right)\nonumber\\
 & & \prod_{i,I,J}\Gamma_{e}\left(\sqrt{pq}\beta_{1}\beta_{5}\gamma_{3}\gamma_{5}^{-1}u_{2,3}^{(i)}\left(u_{1,4,2}^{(J)}\right)^{-1}\right)\Gamma_{e}\left(\beta_{1}^{-2}\beta_{5}^{-2}u_{1,4,2}^{(I)}\left(u_{2,4,2}^{(J)}\right)^{-1}\right)\nonumber\\
 & & \prod_{i,I,J}\Gamma_{e}\left(\sqrt{pq}\beta_{1}\beta_{5}^{-1}\gamma_{3}\gamma_{5}u_{2,3}^{(i)}\left(u_{1,4,1}^{(J)}\right)^{-1}\right)\Gamma_{e}\left(\beta_{1}^{-2}\beta_{5}^{2}u_{1,4,1}^{(I)}\left(u_{2,4,1}^{(J)}\right)^{-1}\right)\nonumber\\
 & & \times\left(u_{1}\to u_{2},u_{2}\to u_{3},\beta_{1}\to\beta_{2},\beta_{2}\to\beta_{3},\beta_{3}\to\beta_{1},\beta_{4}\to\beta_{4}^{-1},\beta_{5}\to\beta_{5}^{-1}\right)\nonumber\\
 & & \left.\times\left(u_{1}\to u_{3},u_{2}\to v_{1},\beta_{1}\to\beta_{3},\beta_{2}\to\beta_{1},\beta_{3}\to\beta_{2}\right)\right]\nonumber\\
 & & \nonumber\\
 & & \times\left(u\leftrightarrow v,\beta_{4}\to\beta_{4}^{-1},\beta_{5}\to\beta_{5}^{-1}\right)\,,
\ee
where the multiplications of small letters $i,j$ runs from $1$ to $2N$ and for capital letters $I,J$ runs from $1$ to $N$.
The multiplication with the assignment brackets indicates multiplication by the same terms appearing in the same square bracket differing by the indicated assignments. In total each square bracket should have multiplications of three copies of the same expression only differing by the written assignments. The last assignment bracket indicates multiplication by the entire expression with the new assignments.

We initiate the flow with the baryonic vev setting $(\sqrt{pq} \beta_{2}\gamma_{2}^{-1})^{2N}=1$, to implement it we define $\gamma_{2}=\left(pq\right)^{1/4}\epsilon^{-1}$ and $\beta_{2}=\left(pq\right)^{-1/4}\epsilon^{-1}$. This flow Higgses the $SU(2N)_{u_{1,3}}$ and $SU(2N)_{v_{1,3}}$ gauge symmetries, and makes some of the fields massive. These massive fields are part of singlet operators that couple to flip fields; thus these flip fields decouple in the IR as well. 

The resulting theory is identified with four fundamental flux tubes like the ones used in the first place to build the torus, but with $p=1$ and another two unidentified building blocks glued together. These can be divided to two blocks of two flux tubes and on unidentified building block. We identify these fundamental building blocks such that the flux tubes are $\Phi$-glued to one another and the new building block, which we will identify as the trinion of the $p=1$ case is $\Phi$-glued from one side and $S$-glued from the other. This is done in a very similar manner to the derivation in \cite{Razamat:2019ukg}, and due to the complexity introduced by considering the non minimal case we will not display the full index of the torus after the flow.

From the above procedure we identify the trinion with two maximal punctures and minimal puncture and its index is given by
\be
\mathcal{I}_{T}^{N,p=1} & = & \left(\frac{\kappa^{2N-1}}{(2N)!}\right)^{2}\oint\frac{du_{2,2}^{(i)}}{2\pi iu_{2,2}^{(i)}}\oint\frac{du_{3,3}^{(i)}}{2\pi iu_{3,3}^{(i)}}\frac{1}{\prod_{i\ne j}\Gamma_{e}\left(u_{2,2}^{(i)}\left(u_{2,2}^{(j)}\right)^{-1}\right)\Gamma_{e}\left(u_{3,3}^{(i)}\left(u_{3,3}^{(j)}\right)^{-1}\right)}\nonumber\\
 & & \Gamma_{e}\left(\left(pq\right)^{\frac{2-N}{2}}\frac{\beta_{4}^{\pm2N}}{\epsilon^{2N}}\right)\Gamma_{e}\left(\left(pq\right)^{\frac{2-N}{2}}\frac{\beta_{1}^{2N}}{\epsilon^{2N}}\right)\Gamma_{e}\left(\left(pq\right)^{\frac{2-N}{2}}\frac{\beta_{3}^{2N}}{\epsilon^{2N}}\right)\Gamma_{e}\left(\left(pq\right)^{\frac{2-N}{2}}\frac{\beta_{5}^{\pm2N}}{\epsilon^{2N}}\right)\nonumber\\
 & & \Gamma_{e}\left(\left(pq\right)^{1/2}\epsilon^{2}\beta_{4}^{2}u_{2,1,1}^{(I)}\left(u_{3,1,1}^{(J)}\right)^{-1}\right)\Gamma_{e}\left(\left(pq\right)^{1/2}\epsilon^{2}\beta_{4}^{-2}u_{2,1,2}^{(I)}\left(u_{3,1,2}^{(J)}\right)^{-1}\right)\nonumber\\
 & & \Gamma_{e}\left(\left(pq\right)^{1/4}\epsilon^{-1}\beta_{4}^{-1}\gamma_{1}\gamma_{4}u_{3,1,1}^{(I)}\left(u_{2,2}^{(j)}\right)^{-1}\right)\Gamma_{e}\left(\left(pq\right)^{1/4}\epsilon^{-1}\beta_{4}\gamma_{1}\gamma_{4}^{-1}u_{3,1,2}^{(I)}\left(u_{2,2}^{(j)}\right)^{-1}\right)\nonumber\\
 & & \Gamma_{e}\left(\left(pq\right)^{1/4}\epsilon^{-1}\beta_{4}^{-1}\gamma_{1}^{-1}\gamma_{4}^{-1}\left(u_{2,1,1}^{(I)}\right)^{-1}u_{2,2}^{(j)}\right)\Gamma_{e}\left(\left(pq\right)^{1/4}\epsilon^{-1}\beta_{4}\gamma_{1}^{-1}\gamma_{4}\left(u_{2,1,2}^{(I)}\right)^{-1}u_{2,2}^{(j)}\right)\nonumber\\
 & & \Gamma_{e}\left(\left(pq\right)^{1/4}\epsilon\beta_{3}\left(u_{2,2}^{(i)}\right)^{-1}u_{2,3}^{(j)}\right)\Gamma_{e}\left(\left(pq\right)^{1/4}\epsilon\beta_{1}^{-1}u_{2,3}^{(i)}\left(u_{3,3}^{(j)}\right)^{-1}\right)\nonumber\\
 & & \Gamma_{e}\left(\sqrt{pq}\epsilon^{-2}u_{3,2}^{(i)}\left(u_{2,3}^{(j)}\right)^{-1}\right)\nonumber\\
 & & \Gamma_{e}\left(\left(pq\right)^{1/4}\epsilon\beta_{1}\left(u_{3,2}^{(i)}\right)^{-1}u_{3,3}^{(j)}\right)\Gamma_{e}\left(\left(pq\right)^{1/4}\epsilon\beta_{3}^{-1}u_{2,2}^{(i)}\left(u_{3,2}^{(j)}\right)^{-1}\right)\nonumber\\
 & & \Gamma_{e}\left(\left(pq\right)^{1/4}\epsilon^{-1}\beta_{5}^{-1}\gamma_{3}\gamma_{5}^{-1}u_{3,3}^{(i)}\left(u_{2,4,2}^{(J)}\right)^{-1}\right)\Gamma_{e}\left(\left(pq\right)^{1/4}\epsilon^{-1}\beta_{5}\gamma_{3}\gamma_{5}u_{3,3}^{(i)}\left(u_{2,4,1}^{(J)}\right)^{-1}\right)\nonumber\\
 & & \Gamma_{e}\left(\left(pq\right)^{1/4}\epsilon^{-1}\beta_{5}\gamma_{3}^{-1}\gamma_{5}^{-1}\left(u_{3,3}^{(i)}\right)^{-1}u_{3,4,1}^{(J)}\right)\Gamma_{e}\left(\left(pq\right)^{1/4}\epsilon^{-1}\beta_{5}^{-1}\gamma_{3}^{-1}\gamma_{5}\left(u_{3,3}^{(i)}\right)^{-1}u_{3,4,2}^{(J)}\right)\nonumber\\
 & & \Gamma_{e}\left(\left(pq\right)^{1/2}\epsilon^{2}\beta_{5}^{2}u_{2,4,2}^{(I)}\left(u_{3,4,2}^{(J)}\right)^{-1}\right)\Gamma_{e}\left(\left(pq\right)^{1/2}\epsilon^{2}\beta_{5}^{-2}u_{2,4,1}^{(I)}\left(u_{3,4,1}^{(J)}\right)^{-1}\right)\nonumber\\
 & & \underline{\Gamma_{e}\left(\left(pq\right)^{1-N/2}\epsilon^{2N}\gamma_{4}^{-2N}\right)\Gamma_{e}\left(\left(pq\right)^{1-N/2}\epsilon^{2N}\beta_{4}^{2N}\right)\Gamma_{e}\left(\left(pq\right)^{1-N/2}\epsilon^{2N}\gamma_{1}^{2N}\right)}\nonumber\\
 & & \underline{\Gamma_{e}\left(\left(pq\right)^{1-N/2}\epsilon^{2N}\gamma_{5}^{2N}\right)\Gamma_{e}\left(\left(pq\right)^{1-N/2}\epsilon^{2N}\beta_{5}^{-2N}\right)\Gamma_{e}\left(\left(pq\right)^{1-N/2}\epsilon^{2N}\gamma_{3}^{-2N}\right)},
\ee
where in the last two lines there are flipping fields added for consistency with the $N=1$ case appearing in \cite{Razamat:2019ukg}. Finally we take $\beta_{i\ge2}^{new}=\beta_{i+1}$ and $\gamma_{i\ge2}^{new}=\gamma_{i+1}$, and find the trinion for $p=1$ in the form of \eqref{E:DTrinion}. This procedure can be generalized to any $N$ by repeating the same steps.

\section*{Acknowledgments}

We are grateful to Shlomo Razamat for useful discussions and comments.
This research is supported by Israel Science Foundation under grant no. 2289/18 and by I-CORE  Program of the Planning and Budgeting Committee, by a Grant No. I-1515-303./2019 from the GIF, the German-Israeli Foundation for Scientific Research and Development, and
by BSF grant no. 2018204.

\vspace{10pt}
\begin{appendix}
\vspace{10pt}

\section{The $\mathcal{N}=1$ superconformal index}\label{A:indexdefinitions}
In this appendix we summarize the background for the $\mathcal{N}=1$ superconformal index, known results and conventions \cite{Kinney:2005ej,Romelsberger:2005eg,Dolan:2008qi}. For a more thorough derivation and definitions see \cite{Rastelli:2016tbz}.
The witten index in radial quantization is defined to be the index of an SCFT. In $4d$ it can be defined as a trace over the Hilbert space of the theory quantized on $\mathbb{S}^3$,
\be
\mathcal{I}\left(\mu_i\right)=Tr(-1)^F e^{-\beta \delta} e^{-\mu_i \mathcal{M}_i},
\ee
where $\delta\triangleq \half \left\{\mathcal{Q},\mathcal{Q}^{\dagger}\right\}$, with $\mathcal{Q}$ and $\mathcal{Q}^{\dagger}=\mathcal{S}$ one of the Poincar\'e supercharges, and its conjugate conformal supercharge, respectively. $\mathcal{M}_i$ are $\mathcal{Q}$-closed conserved charges and $\mu_i$ their associated chemical potentials. Non-vanishing contributions come from states with $\delta=0$ making the index independent on $\beta$. This is true since supersymmetry imposes that states with $\delta>0$ come in boson/fermion pairs.

For $\mathcal{N}=1$, the supercharges are $\left\{\mathcal{Q}_{\alpha},\,\mathcal{S}^{\alpha} \triangleq \mathcal{Q}^{\dagger\alpha},\,\widetilde{\mathcal{Q}}_{\dot{\alpha}},\,\widetilde{\mathcal{S}}^{\dot{\alpha}} \triangleq \widetilde{\mathcal{Q}}^{\dagger\dot{\alpha}}\right\}$, with $\alpha=\pm$ and $\dot{\alpha}=\dot{\pm}$ the respective $SU(2)_1$ and $SU(2)_2$ indices of the isometry group of $\mathbb{S}^3$ ($Spin(4)=SU(2)_1 \times SU(2)_2$).
Different choices of $\mathcal{Q}$ in the definition of the index lead to physically equivalent indices; thus, we can choose for example $\mathcal{Q}=\widetilde{\mathcal{Q}}_{\dot{-}}$. This choice leads to the following index formula,
\be
\mathcal{I}\left(p,q\right)=Tr(-1)^F p^{j_1 + j_2 +\half r} q^{j_2 - j_1 +\half r}.
\ee
where $r$ is the generator of the $U(1)_r$ R-symmetry, and $j_1$ and $j_2$ are the Cartan generators of $SU(2)_1$ and $SU(2)_2$, respectively.

To compute the index we list all the gauge invariant operators we can construct from field modes. The modes are conventionally called "letters" while the operators are called "words". The single-letter index for a vector multiplet and a chiral multiplet transforming in the representation $\mathcal{R}$ of the gauge and flavor group is,
\be
i_V \left(p,q,U\right) & = & \frac{2pq-p-q}{(1-p)(1-q)} \chi_{adj}\left(U\right), \nonumber\\
i_{\chi(r)}\left(p,q,U,V\right) & = & 
\frac{(pq)^{\half r} \chi_{\mathcal{R}} \left(U,V\right) - (pq)^{\frac{2-r}{2}} \chi_{\overline{\mathcal{R}}} \left(U,V\right)}{(1-p)(1-q)},
\ee
where $\chi_{\mathcal{R}} \left(U,V\right)$ denote the characters of $\mathcal{R}$ and $\chi_{\overline{\mathcal{R}}} \left(U,V\right)$ denote the characters of the conjugate representation $\overline{\mathcal{R}}$, with $U$ the gauge group matrix and $V$ the flavor group matrix.

Now we can use the single letter indices to write the full index by listing all the words and projecting them to gauge invariants by integrating over the Haar measure of the gauge group. This takes the general form
\be
\mathcal{I} \left(p,q,V\right)=\int \left[dU\right] \prod_{k} PE\left[i_k\left(p,q,U,V\right)\right],
\ee
where $PE[i_k]$ is the plethystic exponent of the single-letter index of the $k$-th multiplet, listing all the words, and $k$ labels the different multiplets. The plethystic exponent is given by
\be
PE\left[i_k\left(p,q,U,V\right)\right] \triangleq \exp \left\{ \sum_{n=1}^{\infty} \frac{1}{n} i_k\left(p^n,q^n,U^n,V^n\right) \right\}.
\ee

Focusing on the case of $SU(N)$ gauge group relevant for this paper. The full contribution of a chiral superfield in the fundamental representation of $SU(N)$ with R-charge $r$ can be written in terms of elliptic gamma functions $\Gamma_e(z)$, as follows
\be
PE\left[i_k\left(p,q,U\right)\right] & \equiv & \prod_{i=1}^{N} \Gamma_e \left((pq)^{\half r} z_i \right), \nonumber \\
\Gamma_e(z)\triangleq\Gamma\left(z;p,q\right) & \equiv & \prod_{n,m=0}^{\infty} \frac{1-p^{n+1} q^{m+1}/z}{1-p^n q^m z},
\ee
where $\{z_i\}$ with $i=1,...,N$ are the fugacities parameterizing the Cartan subalgebra of $SU(N)$, with $\prod_{i=1}^{N} z_i = 1$. In addition, it is common to use the shorten notation
\be
\Gamma_e \left(u z^{\pm n} \right)=\Gamma_e \left(u z^{n} \right)\Gamma_e \left(u z^{-n} \right).
\ee

In a similar manner we can write the full contribution of the vector multiplet transforming in the adjoint representation of $SU(N)$, together with the matching Haar measure and projection to gauge invariants as
\be
\frac{\kappa^{N-1}}{N !} \oint_{\mathbb{T}^{N-1}} \prod_{i=1}^{N-1} \frac{dz_i}{2\pi i z_i} \prod_{k\ne \ell} \frac1{\Gamma_e(z_k/z_\ell)}\cdots,
\ee
where the dots denote that it will be used in addition to the full matter multiplets transforming under the gauge group. The integration is a contour integration over the maximal torus of the gauge group, and $\kappa$ is the index of a $U(1)$ free vector multiplet defined as
\be
\kappa \triangleq (p;p)(q;q),
\ee
where
\be
(a;b) \triangleq \prod_{n=0}^\infty \left( 1-ab^n \right)
\ee
is the q-Pochhammer symbol. 

\section{S-duality proof for exchanging minimal punctures}\label{A:MinPuncDuality}
In this appendix we will prove using Seiberg duality \cite{Seiberg:1994pq} and S-duality graphically on the quiver diagram, that two minimal punctures with $U(1)$ symmetry are interchangeable when S-gluing two $p=1$ Non minimal D-type trinions. We expect similar proofs can be performed for $p>1$, but we will not display such proofs as their complexity increase with $p$. Some indications for the duality under the exchange of two minimal punctures for any $p$, is the fact that all 't Hooft anomalies related to the punctures match, and that indices match under expansion in fugacities. In addition, a similar proof can be performed in the case of $\Phi$-gluing two trinions. In the presented proof we will not show any of the flip fields of the two glued trinions as one can see that they are symmetric under the exchange of minimal punctures from the get go. These include the $B_i$ fields remaining after the gluing as they are independent on the minimal punctures fugacities.

\begin{figure}[t]
	\centering
  	\includegraphics[scale=0.37]{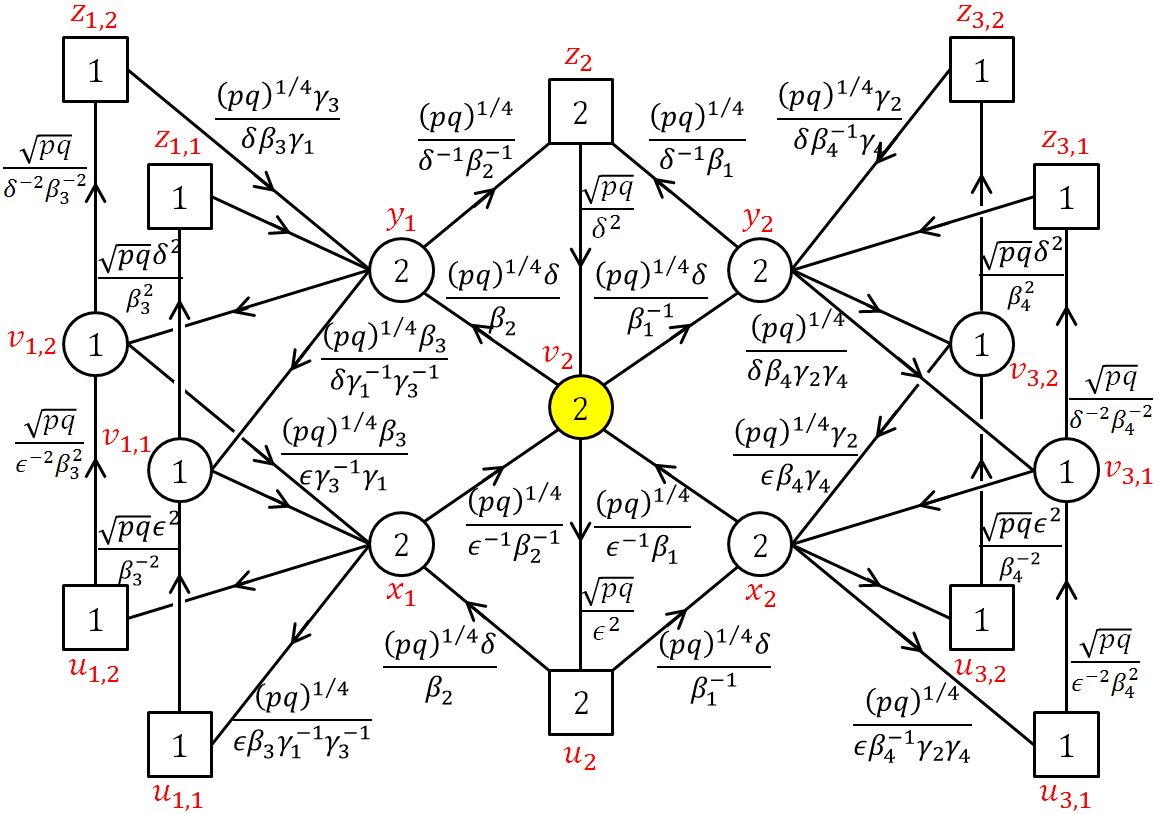}
    \caption{A quiver diagram of two Non minimal D-type trinions with $p=1$ $S$-glued together. In this diagram we use Seiberg duality on the $SU(2N)_{v_2}$ gauge node (yellow).}
    \label{F:2MinExchange1}
\end{figure}

\begin{figure}[t]
	\centering
	\includegraphics[scale=0.37]{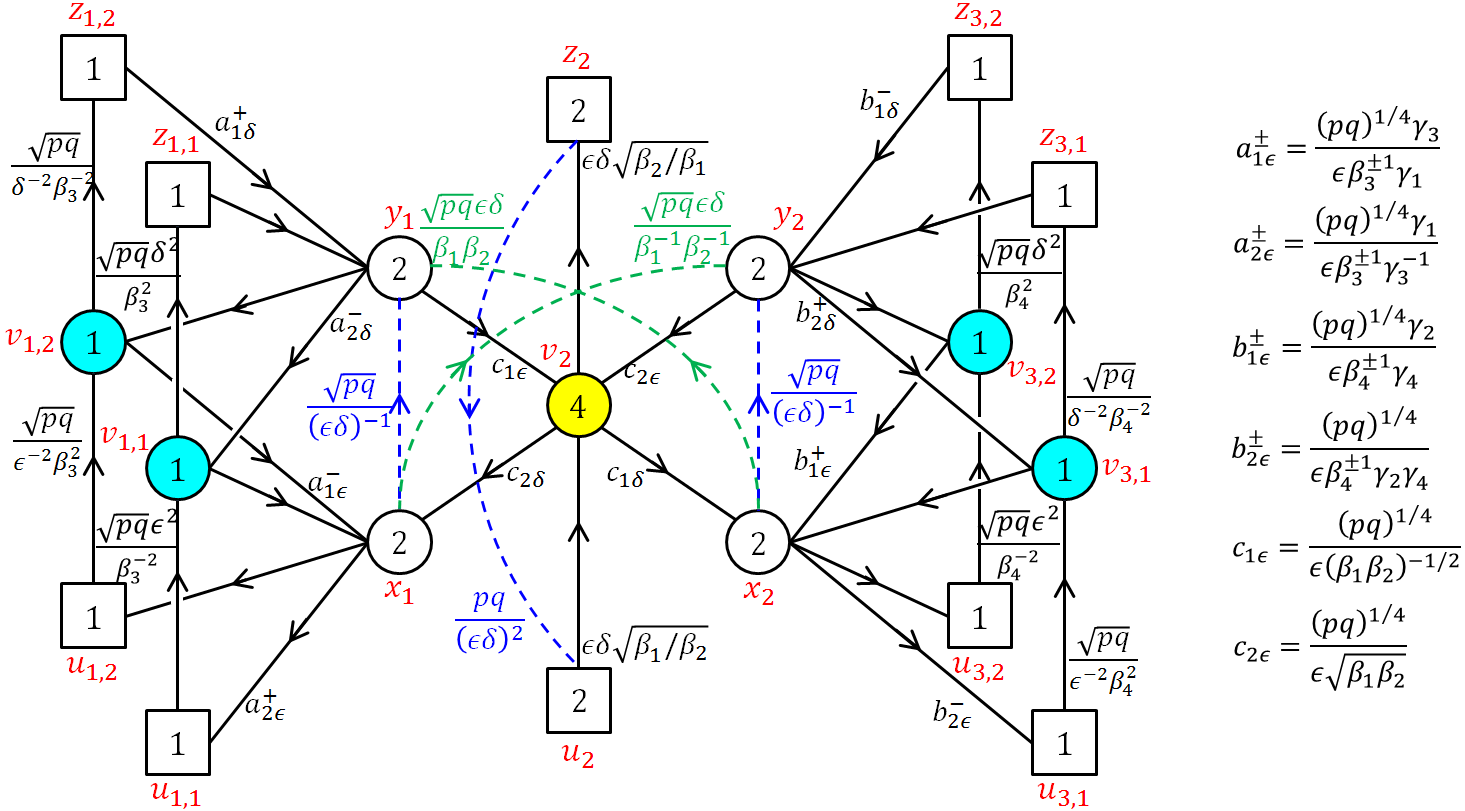}
    \caption{A quiver diagram of two Non minimal D-type trinions with $p=1$ $S$-glued together after the first Seiberg duality. On this quiver we perform Seiberg duality on the four $SU(N)_{v_{i,j}}$ symmetries (light blue). Notice that some of the fields are defined on the right for clarity.}
    \label{F:2MinExchange2}
\end{figure}

After these preliminaries we can get to the proof itself. Starting from two $S$-glued trinions appearing on Figure \ref{F:2MinExchange1}. We preform the first Seiberg duality on the middle $SU(2N)_{v_2}$ gauge node which has $6N$ flavors. The resulting quiver is shown in Figure \ref{F:2MinExchange2}, where the $SU(2N)_{v_2}$ gauge node is replaced with an $SU(4N)$ gauge node.

\begin{figure}[t]
	\centering
  	\includegraphics[scale=0.37]{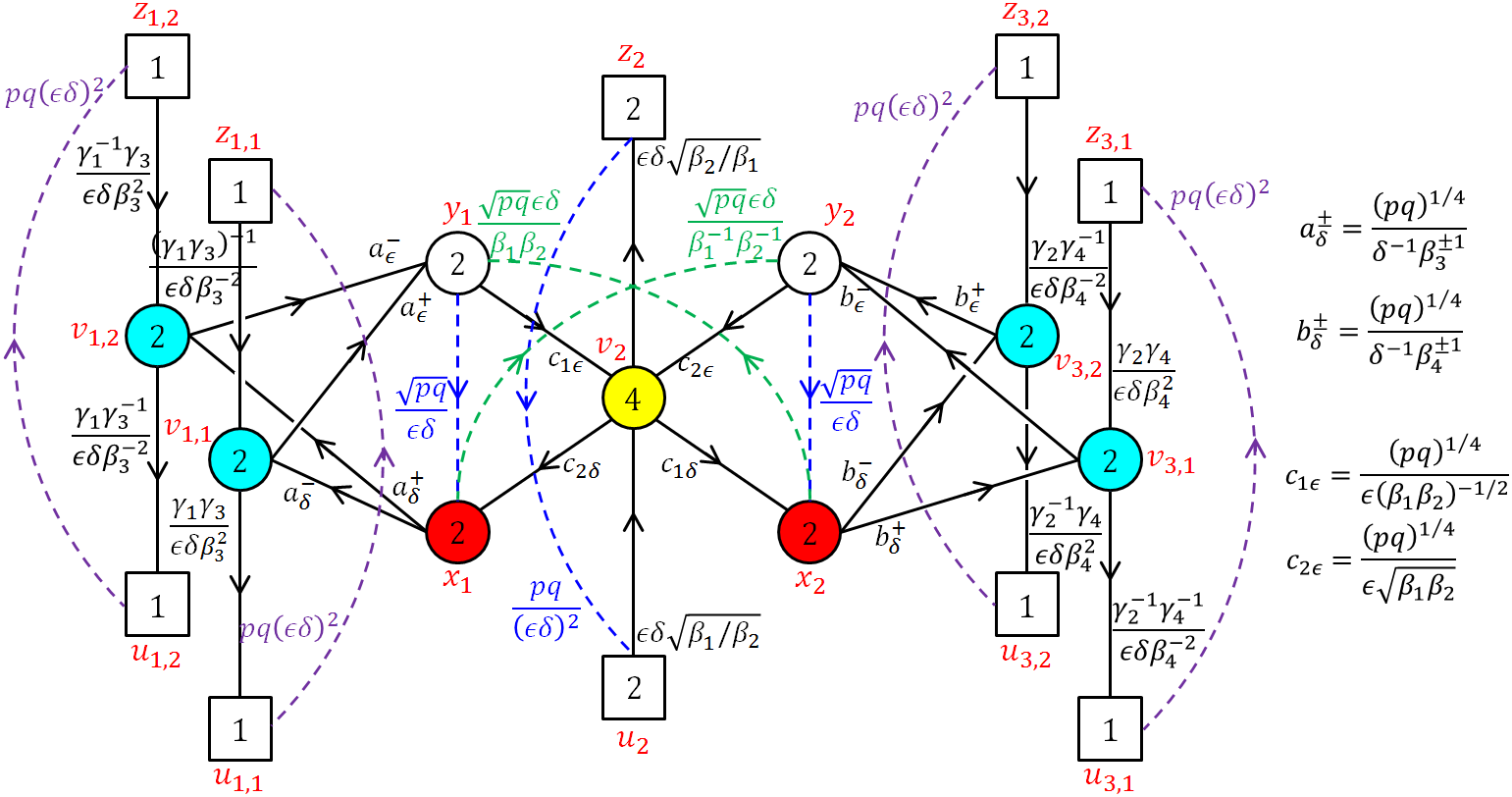}
    \caption{A quiver diagram of two Non minimal D-type trinions with $p=1$ $S$-glued together after five Seiberg dualities. On this quiver we perform Seiberg duality on the two $SU(2N)_{x_{i}}$ symmetries (red). Notice that some of the fields are defined on the right for clarity.}
    \label{F:2MinExchange3}
\end{figure}

Next, we perform four additional Seiberg dualities on the gauge nodes $SU(N)_{v_{1,1}}$, $SU(N)_{v_{1,2}}$, $SU(N)_{v_{3,1}}$, and $SU(N)_{v_{3,2}}$ all with $3N$ flavors. In the resulting quiver these $SU(N)$ nodes get replaced with $SU(2N)$ gauge nodes, see Figure \ref{F:2MinExchange3}.

\begin{figure}[t]
	\centering
  	\includegraphics[scale=0.37]{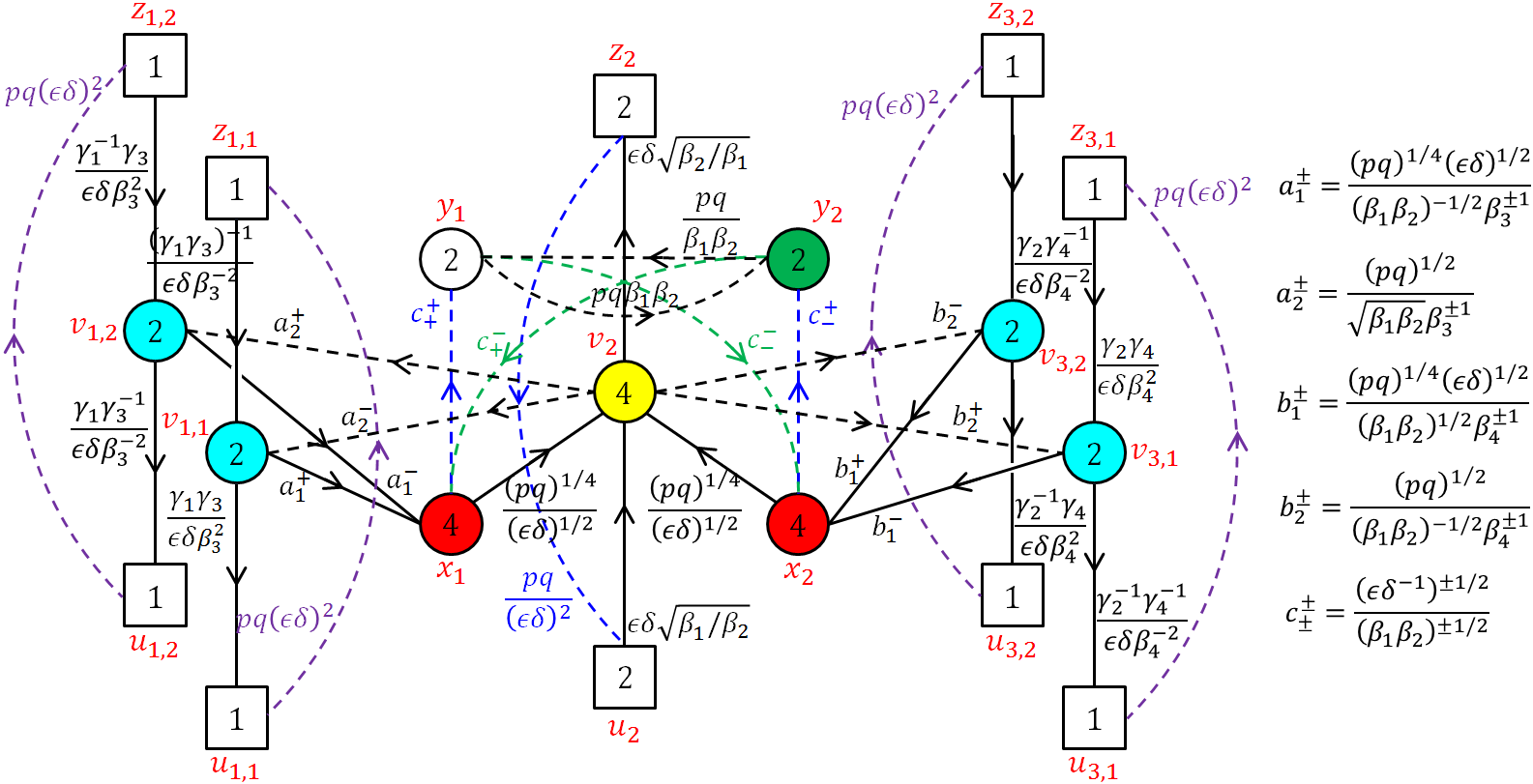}
    \caption{A quiver diagram of two Non minimal D-type trinions with $p=1$ $S$-glued together after seven Seiberg dualities. On this quiver we perform Seiberg duality on the $SU(2N)_{y_{2}}$ gauge node (green). Notice that some of the fields are defined on the right for clarity.}
    \label{F:2MinExchange4}
\end{figure}

The next step is to perform two more Seiberg dualities on the $SU(2N)_{x_1}$ and $SU(2N)_{x_2}$ gauge nodes both with $6N$ flavors. In the transformed quiver both $SU(2N)$ gauge nodes get replaced with $SU(4N)$ gauge nodes, see Figure \ref{F:2MinExchange4}.

\begin{figure}[t]
	\centering
  	\includegraphics[scale=0.37]{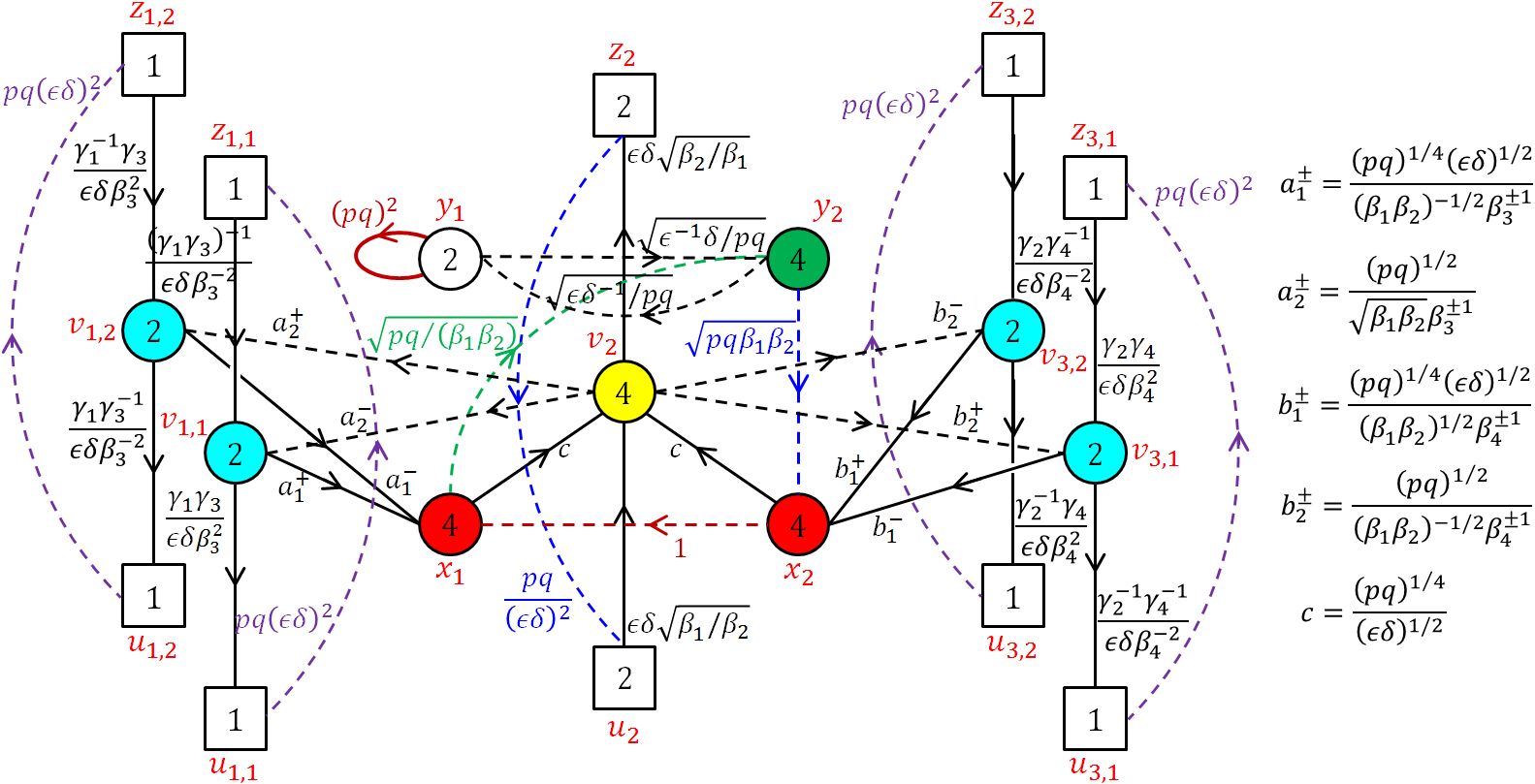}
    \caption{A quiver diagram of two Non minimal D-type trinions with $p=1$ $S$-glued together after eight Seiberg dualities. On this quiver we perform S-duality on the $SU(2N)_{y_{1}}$ gauge symmetry. Notice that some of the fields are defined on the right for clarity.}
    \label{F:2MinExchange5}
\end{figure}

The final Seiberg duality we employ is on the $SU(2N)_{y_2}$ gauge node with $6N$ flavors. The resulting quiver appears on Figure \ref{F:2MinExchange5} with the $SU(2N)_{y_2}$ gauge exchanged with an $SU(4N)$ gauge node.

After all these Seiberg dualities we find a quiver diagram symmetric under the exchange of $\delta$ and $\epsilon$ except for the fundamental and antifundamental fields that transform under the $SU(2N)_{y_1}$ gauge symmetry. This $SU(2N)$ gauge node has one adjoint and $4N$ fundamental and antifundamental fields; therefore, we can use S-duality on it. The S-dual frame exchanges the fundamental fields with the antifundamental fields. The resulting quiver diagram is the same as the one before this S-duality only with $\delta$ and $\epsilon$ exchanged. At last, we can use the same Seiberg dualities mentioned above in reverse to get back to the quiver original form only with $\delta$ and $\epsilon$ exchanged. This proves that the $U(1)$ minimal punctures obey S-duality and the two quivers with the two punctures exchanged are indeed dual to one another as required.

\section{The $D_{p+3}$ 't Hooft anomaly predictions from $6d$}
\label{A:Anomalies}
Here we will develop the $4d$ anomaly polynomial by reducing the $6d$ anomaly polynomial on a Riemann surface with fluxes. The $6d$ anomaly polynomial was given in \cite{Ohmori:2014kda}, and we reproduce it here
\be
\label{E:Dtype6dAP}
I_{8}^{D_{p+3}} & = & \frac{1}{24}\left(16N^{3}(p+1)^{2}-2N\left(4p^{2}+20p+15\right)+2p^{2}+11p+14\right)C_{2}^{2}\left(R\right)\nonumber\\
 & & +\frac{1}{48}\left(-2N\left(2p^{2}+10p+7\right)+2p^{2}+11p+14\right)C_{2}\left(R\right)p_{1}\left(T\right)\nonumber\\
 & & +\frac{1}{2}\left(-2N(p+1)+p+2\right)\left(C_{2}\left(SO(2p+6)_{\beta}\right)_{V}+C_{2}\left(SO(2p+6)_{\gamma}\right)_{V}\right)C_{2}\left(R\right)\nonumber\\
 & & +\frac{\left(p+2\right)}{24}\left(C_{2}\left(SO(2p+6)_{\beta}\right)_{V}+C_{2}\left(SO(2p+6)_{\gamma}\right)_{V}\right)p_{1}\left(T\right)\nonumber\\
 & & +\frac{\left(2N(p+2)-3\right)}{24N}\left(C_{2}^{2}\left(SO(2p+6)_{\beta}\right)_{V}+C_{2}^{2}\left(SO(2p+6)_{\gamma}\right)_{V}\right)\nonumber\\
 & & +\frac{1}{4N}C_{2}\left(SO(2p+6)_{\beta}\right)_{V}C_{2}\left(SO(2p+6)_{\gamma}\right)_{V}\nonumber\\
 & & -\frac{\left(p-1\right)}{6}\left(C_{4}\left(SO(2p+6)_{\beta}\right)_{V}+C_{4}\left(SO(2p+6)_{\gamma}\right)_{V}\right)\nonumber\\
 & & +\frac{\left(30N+7p^{2}+77p+82\right)p_{1}\left(T\right)^{2}-4\left(30N+2p^{2}+11p-14\right)p_{2}\left(T\right)}{5760}\,.
\ee
where $C_{i}(G)_{\boldsymbol{R}}$ is the $i$-th Chern class of the global symmetry $G$, evaluated in the representation $\boldsymbol{R}$ ($V$ stands for the vector representation), $C_{2}(R)$ stands for the second Chern class of the $SU(2)_{R}$ six dimensional R-symmetry in the fundamental representation. In addition, $p_{1}(T)$ and $p_{2}(T)$ are the first and second Pontryagin classes, respectively.

We want to calculate anomalies for a general flux compactification; therefore we will decompose both $SO(2p+6)$ groups to their Cartan $U(1)^{p+3}$. For the vector representation the decomposition takes the form
\be
\boldsymbol{V}\left(\beta\right)\to\sum_{i=1}^{p+3}\left(\beta_{i}+\beta_{i}^{-1}\right)\,,
\ee
where $\beta_{i}$ are the fugacities for the chosen Cartans. This decomposition translates to the following Chern classes decomposition
\be
C_{2}\left(SO(2p+6)_{\beta}\right)_{V}&\to&-\sum_{i=1}^{p+3}C_{1}^{2}\left(U(1)_{\beta_{i}}\right)\,,\nonumber\\
C_{4}\left(SO(2p+6)_{\beta}\right)_{V}&\to&-\frac{1}{2}\sum_{i=1}^{p+3}C_{1}^{4}\left(U(1)_{\beta_{i}}\right)+\frac{1}{2}\sum_{i,j=1}^{p+3}C_{1}^{2}\left(U(1)_{\beta_{i}}\right)C_{1}^{2}\left(U(1)_{\beta_{j}}\right)\,.
\ee
The exact same decompositions hold for the second $SO(2p+6)_{\gamma}$ by replacing $\beta$ with $\gamma$.

The next step after decomposing the above groups to their Cartans is the compactification itself. We want to compactify the $6d$ anomaly polynomial eight-form on a Riemann surface $\Sigma$ of genus $g$ and a general flux.\footnote{By general flux we mean we will take an integer non vanishing flux to all the Cartan symmetries, but some of these can later be set to vanish.} The flux setting is done by taking $\int_{\Sigma}C_{1}\left(U(1)_{\beta_{i}}\right)=-N_{b_{i}}$ and $\int_{\Sigma}C_{1}\left(U(1)_{\gamma_{j}}\right)=-N_{c_{j}}$, where $N_{b_{i}}$ and $N_{c_{j}}$ are integers. The R-symmetry inherited from $6d$ under the embedding $U(1)_{R}\subset SU(2)_{R}$\footnote{This is not necessarily the superconformal R-symmetry.} does not necessarily preserve supersymmetry. This can be fixed by twisting the $SO(2)$ acting on the tangent space of the Riemann surface with the Cartan of $SU(2)_{R}$, leading to the Chern class decomposition $C_{2}\left(R\right)\to-C_{1}\left(R'\right)^{2}+2(1-g)tC_{1}\left(R'\right)+\mathcal{O}\left(t^{2}\right)$. The final step before the compactification is to set
\be
C_{1}\left(U(1)_{\beta_{i}}\right)=-N_{b_{i}}t+\epsilon_{\beta_{i}}C_{1}\left(R'\right)+2NC_{1}\left(U(1)_{F_{\beta_{i}}}\right)\,.
\ee
The first term is required to set the flux to be $N_{b_{i}}$, where $t$ is a unit flux two form on $\Sigma$, meaning we set $\int_{\Sigma}t=1$. The second term is required due to possible mixing of flavor $U(1)$ symmetries with the R-symmetry to generate the superconformal R-symmetry, where the mixing parameters $\epsilon_{\beta_{i}}$ will be determined by $a$-maximization \cite{Intriligator:2003jj}. The last term denotes the $4d$ curvature of the chosen U(1). The same needs to be done for the Cartans denoted by $\gamma$ with the matching flux.

The final step is the compactification itself, where we first plug all the above replacements to the $6d$ anomaly polynomial given in \eqref{E:Dtype6dAP}, and then compactify by integrating over the Riemann surface $\Sigma$. We find
\be
I_{6}^{D_{p+3}} & = & \frac{1}{6}\left(16N^{3}\left(p+1\right)^{2}-2N\left(4p^{2}+20p+15\right)+2p^{2}+11p+14\right)\left(g-1\right)C_{1}^{3}\left(R\right)\nonumber\\
 & & +\frac{1}{24}\left(2N\left(2p^{2}+10p+7\right)-2p^{2}-11p-14\right)\left(g-1\right)C_{1}\left(R\right)p_{1}\left(T_{4}\right)\nonumber\\
 & & -4N^{2}\left(2N(p+1)-p-2\right)\left(g-1\right)C_{1}\left(R\right)\sum_{i=1}^{p+3}\left(C_{1}^{2}\left(\beta_{i}\right)+C_{1}^{2}\left(\gamma_{i}\right)\right)\nonumber\\
 & & +2N\left(2N(p+1)-p-2\right)\left(g-1\right)C_{1}^{2}\left(R\right)\sum_{i=1}^{p+3}\left(N_{b_{i}}C_{1}\left(\beta_{i}\right)+N_{c_{i}}C_{1}\left(\gamma_{i}\right)\right)\nonumber\\
 & & +\frac{N\left(p+2\right)}{6}p_{1}\left(T_{4}\right)\sum_{i=1}^{p+3}\left(N_{b_{i}}C_{1}\left(\beta_{i}\right)+N_{c_{i}}C_{1}\left(\gamma_{i}\right)\right)\nonumber\\
 & & -\frac{8N^{3}\left(p-1\right)}{3}\sum_{i=1}^{p+3}\left(N_{b_{i}}C_{1}^{3}\left(\beta_{i}\right)+N_{c_{i}}C_{1}^{3}\left(\gamma_{i}\right)\right)\nonumber\\
 & & -4N^{2}\sum_{i,j=1}^{p+3}N_{b_{i}}C_{1}\left(\beta_{i}\right)\left(\left(2N-1\right)C_{1}^{2}\left(\beta_{j}\right)+C_{1}^{2}\left(\gamma_{j}\right)\right)\nonumber\\
 & & -4N^{2}\sum_{i,j=1}^{p+3}N_{c_{i}}C_{1}\left(\gamma_{i}\right)\left(C_{1}^{2}\left(\beta_{j}\right)+\left(2N-1\right)C_{1}^{2}\left(\gamma_{j}\right)\right)\,,
\ee
where the chosen R-charge is the one inherited from $6d$, meaning we take $\epsilon_{\beta_{i}}=\epsilon_{\gamma_{i}}=0$ for all $i$. In addition, we replaced $C_{1}\left(U(1)_{F_{\beta_{i}}}\right) \to C_{1}\left(\beta_{i}\right)$ and similarly for $\gamma_j$ to shorten the notation. Finally let us specify explicitly all the $4d$ anomalies derived from the above anomaly polynomial for ease of use,
\be
&& Tr\left(U(1)_{R}^{3}\right)=\left(16N^{3}\left(p+1\right)^{2}-2N\left(4p^{2}+20p+15\right)+2p^{2}+11p+14\right)\left(g-1\right)\,,\nonumber\\
&& Tr\left(U(1)_{R}\right)=-\left(2N\left(2p^{2}+10p+7\right)-2p^{2}-11p-14\right)\left(g-1\right)\,,\nonumber\\
&& Tr\left(U(1)_{\beta_{i}/\gamma_{i}}^{3}\right)=-8N^{2}\left(2N\left(p+2\right)-3\right)N_{b_{i}/c_{i}},\quad Tr\left(U(1)_{\beta_{i}/\gamma_{i}}\right)=-4N\left(p+2\right)N_{b_{i}/c_{i}}\,,\nonumber\\
&& Tr\left(U(1)_{R}U(1)_{\beta_{i}/\gamma_{i}}^{2}\right)=-8N^{2}\left(2N\left(p+1\right)-p-2\right)\left(g-1\right)\,,\nonumber\\
&& Tr\left(U(1)_{R}^{2}U(1)_{\beta_{i}/\gamma_{i}}\right)=4N\left(2N\left(p+1\right)-p-2\right)N_{b_{i}/c_{i}}\,,\nonumber\\
&& Tr\left(U(1)_{\beta_{i}/\gamma_{i}}U(1)_{\beta_{j}/\gamma_{j}}^{2}\right)=-8N^{2}\left(2N-1\right)N_{b_{i}/c_{j}}\,,\nonumber\\
&& Tr\left(U(1)_{\beta_{i}/\gamma_{i}}U(1)_{\gamma_{j}/\beta_{j}}^{2}\right)=-8N^{2}N_{b_{i}/c_{j}}\,,
\ee
where the slashes appearing in some of the formulas are correlated, and the anomalies not written vanish.

\end{appendix}


\bibliographystyle{ytphys}
\bibliography{refs}

\end{document}